\documentclass[prb,twocolumn,notitlepage,superscriptaddress]{revtex4-2}
\usepackage{tabularx,graphicx,amsmath,amssymb,wasysym,latexsym,yfonts,pifont,bbm,url}
\UseRawInputEncoding
\usepackage{bm,psfrag,amsmath,amssymb,grffile,float,bbm}
\usepackage{tabularx,graphicx,color}
\usepackage{latexsym}
\usepackage{amssymb}
\usepackage{amsmath}
\usepackage{amsfonts}
\usepackage{upgreek}
\usepackage{bm}
\usepackage{multirow}
\usepackage{enumitem}
\usepackage{overpic}
\usepackage[hidelinks,colorlinks,linkcolor=blue,
citecolor=blue,urlcolor=blue]{hyperref}
\usepackage{graphicx,siunitx}
\usepackage[dvipsnames]{xcolor}
\usepackage{pifont}

\usepackage[normalem]{ulem}
\usepackage{comment}

\usepackage[english]{babel}
\makeatletter
\addto\captionsenglish{}
\addto\extrasenglish{}
\@ifundefined{captionsen}{\adddialect\l@en\l@english}{}
\makeatother

\usepackage{xcolor}
\usepackage{overpic}

\newcommand{\be}{\begin{equation}}
\newcommand{\ee}{\end{equation}}
\newcommand{\bea}{\begin{equation} \begin{aligned}}
\newcommand{\eea}{\end{aligned} \end{equation} }
\newcommand{\bi}{\begin{itemize}}
\newcommand{\ei}{\end{itemize}}

\newcommand{\la}{\lambda}
\renewcommand{\be}{\beta}
\newcommand{\al}{\alpha}
\newcommand{\bpm}{\begin{pmatrix}}
\newcommand{\epm}{\end{pmatrix}}

\renewcommand{\th}{\theta}

\newcommand{\lp}{\left(}
\newcommand{\rp}{\right)}

\newcommand{\del}{\partial}

\newcommand{\Tr}{\text{Tr} \ }

\usepackage{amsmath,amsfonts,amssymb,amsthm,array,mathrsfs}
\usepackage{dsfont}
\usepackage{slashed}
\usepackage{graphics}
\usepackage{float}
\usepackage{verbatim}
\usepackage{color}
\usepackage{tabularx}
\usepackage{mathtools}
\usepackage{subcaption}

\newcommand{\bsl}[1]{\mathbf{#1}} 
\newcommand{\mbf}[1]{\boldsymbol{#1}} 

\newcommand{\ket}[1]{|#1 \rangle}

\newcommand{\ii}{\mathrm{i}}

\newcommand{\dsR}{\mathbb{R}}

\newcommand{\rmg}{\mathrm{g}}
\renewcommand{\Re}{\mathop{\mathrm{Re}}}

\newcommand{\refcite}[1]{Ref.\,\cite{#1}}

\newcommand{\mat}[1]{\left(\begin{matrix}#1\end{matrix}\right)}

\newcommand{\eq}[1]{\begin{equation} #1 \end{equation}}

\newcommand{\eqa}[1]{\begin{align}\begin{split} #1 \end{split}\end{align}}

\usepackage{environ}
\NewEnviron{eqs}{%
\begin{equation}\begin{split}
    \BODY
\end{split}\end{equation}
}

\let\oldAA\AA
\renewcommand{\AA}{\text{\normalfont\oldAA}}

\newcommand{\ie}{{\emph{i.e.}}}

\newcommand{\G}{\mathcal{G}}

\renewcommand{\P}{\mathcal{P}}
\newcommand{\PT}{\mathcal{PT}}
\newcommand{\T}{\mathcal{T}}

\usepackage{cleveref}
\crefname{appendix}{App.}{Apps.}
\crefname{equation}{Eq.}{Eqs.}
\crefname{figure}{Fig.}{Figs.}
\crefname{table}{Tab.}{Tabs.}
\crefname{section}{Sec.}{Secs.}
\creflabelformat{appendix}{[#2#1#3]}

\mathchardef\sGamma="7100
\mathchardef\sDelta="7101
\mathchardef\sTheta="7102
\mathchardef\sLambda="7103
\mathchardef\sXi="7104
\mathchardef\sPi="7105
\mathchardef\sSigma="7106
\mathchardef\sUpsilon="7107
\mathchardef\sPhi="7108
\mathchardef\sPsi="7109
\mathchardef\sOmega="710A
\def\Br{{\boldsymbol{r}}}
\def\BR{{\boldsymbol{R}}}
\def\Bq{{\boldsymbol{q}}}
\def\Bp{{\boldsymbol{p}}}
\def\Bk{{\boldsymbol{k}}}
\def\Bd{{\boldsymbol{d}}}
\def\Bu{{\boldsymbol{u}}}
\def\Bv{{\boldsymbol{v}}}
\def\BG{{\boldsymbol{G}}}
\def\BQ{{\boldsymbol{Q}}}
\def\BDelta{{\boldsymbol{\Delta}}}
\def\Ba{{\boldsymbol{a}}}
\def\Bb{{\boldsymbol{b}}}
\def\Bd{{\boldsymbol{d}}}
\def\Btau{{\boldsymbol{\tau}}}
\def\Bnabla{{\boldsymbol{\nabla}}}
\def\CI{{\cal I}}
\def\nd{^{\vphantom{\dagger}}}
\def\ns{^{\vphantom{*}}}
\def\yd{^\dagger}

\def\BDelta{{\boldsymbol{\Delta}}}
\def\pz{\partial}
\def\Half{{1\over 2}}
\def\half{\frac{1}{2}}

\def\hh{\hskip 0.1em}

\newcommand{\eV}{\mathrm{eV}}

\newcommand{\wt}[1]{\widetilde{#1}}

\begin{document}

\title{Geometric Contribution to the Electron-Phonon Coupling Constant: Checkerboard and Hyper-diamond Models}

\author{Thomas Caligiure}
\affiliation{Department of Physics and Quantum Theory Project, University of Florida, Gainesville, Florida 32611, USA}

\author{Daniel P. Arovas}
\affiliation{Department of Physics, University of California at San Diego, La Jolla, California 92093, USA}

\author{Jiabin Yu}
\email{yujiabin@ufl.edu}
\affiliation{Department of Physics and Quantum Theory Project, University of Florida, Gainesville, Florida 32611, USA}

\date{\today}

\begin{abstract}

 Recent theoretical studies on electron-phonon coupling (EPC) - a key ingredient in the emergence of diverse quantum phases such as superconductivity and charge-density waves - have revealed a distinct contribution to its strength arising from the quantum geometry of electronic bands. Here we investigate the quantum geometric contribution, characterized by the quantum metric, to the dimensionless EPC constant in the checkerboard and hyper-diamond models, under the Gaussian approximation. We show that in the checkerboard model, the geometric contribution can be tuned independently while keeping the energetic part fixed, demonstrating the possibility of enhancing EPC strength through quantum geometry. In the hyper-diamond model, the geometric contribution accounts for roughly $50\%$ of the total EPC constant across all spatial dimensions, generalizing the $d=2$ result for graphene to higher dimensions.

\end{abstract}

\maketitle

\section{Introduction \label{Main: Introduction}}

Referring to the momentum dependence of the Bloch states in the Brillouin zone, quantum geometry has been found to be crucial in the manifestation of various quantum behaviors~\cite{rossi_quantum_2021, torma_essay_2023, liu_quantum_2025, jiang_revealing_2025, yu_quantum_2025, verma_quantum_2026, yang_berry_2026, kitamura_quantum_2026, gao_quantum_2026}, such as superfluid weight~\cite{peotta_superfluidity_2015, julku_geometric_2016, liang_band_2017, torma_quantum_2018, hu_geometric_2019, julku_superfluid_2020, xie_topology-bounded_2020, herzog-arbeitman_superfluid_2022, torma_superconductivity_2022, huhtinen_revisiting_2022, bouzerar_hidden_2023, tian_evidence_2023, yu_euler-obstructed_2023, han_quantum_2024, hirobe_anomalous_2025, pai_isoperimetric_2026, peotta_quantum_2025, banerjee_superfluid_2025, tanaka_superfluid_2025, jiang_superfluid_2025, hu_quantum_2025-1, buthenhoff_functional_2025, zeng_superfluid_2025, yu_universal_2025, lamponen_superconductivity_2025, zeng_superfluid_2025-1, sun_geometric_2026}, unconventional superconductivity \cite{kitamura_quantum_2022, bouzerar_hidden_2023, sun_flat-band_2024, kitamura_spin-triplet_2024, yang_superconductivity_2024, daido_quantum_2024, han_exact_2025, liao_detecting_2025, buthenhoff_functional_2025, jahin_enhanced_2025, shavit_quantum_2025, simon_normal-state_2025, lamponen_superconductivity_2025, hiorth_ab_2026, liu_electric_2026}, optical responses~\cite{aebischer_dielectric_2001, martin_electronic_2011, gao_nonreciprocal_2019, holder_consequences_2020, ahn_superconductivity-induced_2021, chen_quantum-geometry-induced_2021, mitscherling_bound_2022, ahn_riemannian_2022, komissarov_quantum_2024, zhu_anomalous_2024, mendez-valderrama_low-energy_2024, postlewaite_quantum_2024, carmichael_probing_2025, gao_quantum_2026, yu_quantum_2025, jankowski_optical_2025, souza_optical_2025, onishi_quantum_2025, mao_low-energy_2025,li_quantum-metric-based_2026}, and entanglement~\cite{tam_corner_2024, wu_corner_2025}.
In this work, we focus on the role of quantum geometry in the electron-phonon coupling (EPC)~\cite{yu_nontrivial_2024, zhu_anomalous_2024, hu_kagome_2025, pellitteri_phonon_2025, hu_quantum_2025-1, hu_quantum_2025, jiang_quantum_2026, agarwal_geometric_2026}.
In particular, \refcite{yu_nontrivial_2024} revealed the significant quantum geometric contribution to the EPC constant $\la$ in graphene and MgB$_2$.
As the EPC constant $\la$ has a strong positive correlation with the superconducting critical temperatures $T_c$~\cite{mcmillan_transition_1968,allen_transition_1975}, the results in \refcite{yu_nontrivial_2024} imply a potential enhancement of superconducting critical temperatures from a strong quantum geometric properties.

In this work, we generalize the results in \refcite{yu_nontrivial_2024} to two models: the 2D checkerboard model \cite{sun_time-reversal_2008, sun_topological_2009, li_magic-angle_2022} and the hyper-diamond model.
To separate the geometric contribution from the energetic one, we adopt the Gaussian approximation proposed in \refcite{yu_nontrivial_2024}.
In both models, we find two contributions to the EPC constant: one geometric and one energetic. 
We find that the checkerboard model allows us to tune the geometric contribution while keeping the energetic contribution constant to the leading order, serving as the first clear example that demonstrates the principle of enhancing EPC strength by increasing the quantum geometric effect.
The hyper-diamond model is a natural generalization of graphene to any arbitrary dimension, and we find the geometric contribution limits to one half of the total EPC constant as the Fermi energy approaches the charge neutrality.

\section{Review of the Gaussian Approximation \label{Main:EPC_and_GA}}

We start by reviewing Gaussian approximation proposed in \cite{yu_nontrivial_2024} for EPC. 
We only outline the key steps here and leave details in \cref{Appendix: Introducing EPC}.
Under the two center approximation, the general tight-binding Hamiltonian, including ion motions, 
can be written as 
\begin{equation}
\begin{split}
H_{\rm full} &= \sum_{\BR,\BR'}\sum_{\Btau,\Btau'} t\ns_{\Btau\Btau'}
(\BDelta_{\BR\BR'}^{\Btau\Btau'})\,c\yd_{\BR+\Btau} c\nd_{\BR'+\Btau'}\\
\BDelta_{\BR\BR'}^{\Btau\Btau'}&=\BR + \Btau + \Bu\ns_{\BR+\Btau} 
- \BR' - \Btau' - \Bu\ns_{\BR'+\Btau'}
\end{split}
\end{equation}
where $\BR$ labels a Bravais lattice unit cell, $\Btau$ is a crystalline basis vector, $t\ns_{\Btau\Btau'}(\Br)$ is the hopping function, $\Bu\ns_{\BR+\Btau}$ is the lattice deformation for the atom at basis position $\Btau$ in unit cell $\BR$, and $\BDelta_{\BR\BR'}^{\Btau\Btau'}$ is the separation between the atoms located at positions $\BR+\Btau$ and $\BR'+\Btau'$, including the lattice deformation contributions. We neglect spin as it will not change the final expression for the EPC constant, as we presume \textsf{SU}(2) spin symmetry, and assume one orbital-per-ion. $c\yd_{\BR+\Btau}$ creates an electron in unit cell $\BR$ at basis position $\Btau$, with $\{c\nd_{\BR+\Btau},c\yd_{\BR'+\Btau'}\}= \delta\ns_{\BR \BR'}\,\delta\ns_{\Btau\Btau'}$\,, and $\Bu\ns_{\BR+\Btau}$ is the displacement of that ion due to lattice deformation.  Assuming the deformations are small in comparison with the unit cell size, the hopping function may be expanded in powers of the separations $\Bu\ns_{\BR+\Btau}-\Bu\ns_{\BR'+\Btau'}$. 
The zeroth-order term yields the tight-binding electron Hamiltonian 
\eqa{
H_{\rm el} = \sum_{\BR,\Btau\atop\BR',\Btau'} t\ns_{\Btau\Btau'}
(\BR+\Btau- \BR'-\Btau') \,c\yd_{\BR+\Btau} c\nd_{\BR'+\Btau'}\ .
\label{Main_EQ:Electron_TBM}
}
The first-order term produces the leading-order EPC,
\begin{equation}
\begin{split}
H_{\rm el-ph}&=\sum_{\BR,\Btau\atop\BR',\Btau'} (\Bu\ns_{\BR+\Btau}-
\Bu\ns_{\BR'+\Btau'})\cdot\BQ^{\Btau\Btau'}_{\BR\BR'} \, 
c\yd_{\BR+\Btau} c\nd_{\BR'+\Btau'}\\
\BQ^{\Btau\Btau'}_{\BR\BR'}&= {\pz t\ns_{\Btau\Btau'}(\Br)\over\pz\Br}
\bigg|\nd_{\Br = \BR+\Btau -\BR' -\Btau'}\ .
\label{Main_EQ:EPC_Hamiltonian}
\end{split}
\end{equation}
Under the Gaussian approximation \cite{yu_nontrivial_2024}, we assume the hopping function has the form
\begin{equation}
t\ns_{\Btau\Btau'}(\Br) = t^0_{\Btau\Btau'}\, \exp\!\lp \half \gamma\ns_{\Btau\Btau'}\,\Br^2\rp\ ,
\label{Main_EQ:Gaussian Approximation}
\end{equation}
where for simplicity we choose $\gamma\ns_{\Btau\Btau'} \equiv \gamma < 0$  to describe the decay of the hopping as the distance $|\Br|$ increases. 
We transform to reciprocal space using the basis transformation rule  $c\yd_{\BR+\Btau} = \frac{1}{\sqrt{N}}\sum_\Bk e^{-\ii \Bk \cdot (\BR + \Btau)} c\yd_{\Bk,\Btau}$ and $u\nd_{\BR+\Btau,i} = \frac{1}{\sqrt{N}} \sum_{\Bq} e^{\ii \Bq \cdot (\BR + \Btau)}\,u\nd_{\Bq,\Btau,i}$, where $i$ is a Cartesian index, we can show 
\begin{align}
H_{\rm el-ph} & = N^{-1/2} \sum_{\Bk_1, \Bk_2} \sum_{\Btau,i}  c\yd_{\Bk_1}
\big[\chi\nd_\Btau f\nd_i(\Bk_2) - f\nd_i(\Bk_1)\chi\nd_\Btau\big] c\nd_{\Bk_2}\nonumber\\
& \hskip 1.5in \times u\yd_{\Bk_2 - \Bk_1, \Btau, i}\ ,
\end{align}
where $c\yd_\Bk\equiv\big(c\yd_{\Bk,\Btau_1},\ldots,c\yd_{\Bk,\Btau_p}\big)$ is a  $p$-component vector of creation operators at wavevector $\Bk$, with $p$ the total number of basis elements. $N$ is the number of unit cells, and  $[\chi_\Btau]_{\Btau_1 \Btau_2} = \delta_{\Btau, \Btau_1} \delta_{\Btau_1, \Btau_2}$ is a projection matrix onto the $\Btau$ sublattice. $f_i(\Bk) = \ii\hh\gamma\,\pz\ns_{k_i} h(\Bk)$, with $\pz\ns_{k_i} \equiv{\pz\over \pz k_i}$, is a matrix which can be expressed in terms of  the electron Hamiltonian 
$[h(\Bk)]_{\Btau \Btau'} = \sum_{\BR} t_{\Btau \Btau'}(\BR + \Btau - \Btau')\, e^{-\ii \Bk \cdot (\BR + \Btau - \Btau')}$. We write $h(\Bk) = \sum_n E_n(\Bk)\hh P_n(\Bk)$ in its spectral  form, where $E_n(\Bk)$ is the dispersion of band $n$ and $P\nd_n(\Bk) = U\nd_n(\Bk)\hh U_n^\dagger(\Bk)$ is the projection operator. As such, we can see that $f\ns_i(\Bk)$ will split into two terms:
\begin{equation}
\begin{split}
f_i^E (\Bk) &= \ii \gamma \sum_n P_n(\Bk)\, {\pz E_n(\Bk) \over \pz k_i}  \\
f_i^{\rm geo}(\Bk) &= \ii \gamma \sum_n E_n(\Bk)\, {\pz P_n(\Bk)\over\pz k_i}\ . 
\label{Main_EQ:f_Splitting}
\end{split}
\end{equation}
We denote $f_i^{E}(\Bk)$ as the energetic contribution, as it depends on 
$\pz_{k_i} E_n(\Bk)$, which vanishes if the bands are nondispersive (\ie, flat). 
$f_i^{\rm geo}(\Bk)$ is the geometric contribution, since it relies on the momentum 
dependence of the projectors $\pz\ns_{k_i} P_n(\Bk)$, which vanishes for $U_n(\Bk)$ 
with no $\Bk$ dependence (up to a gauge choice).  From this, we can derive general 
expressions for the EPC constant $\la$. We can write the bulk EPC constant 
as \cite{mcmillan_transition_1968} 
\eq{
\lambda = {2D(\mu)\over N\hbar}\cdot{\langle\Gamma\rangle\over\langle\omega^2\rangle} \ ,
}
where $D(\mu)$ is the single-particle density of states, $\mu$ is the chemical potential, and 
$\langle\omega^2\rangle$ is McMillan's mean-square phonon frequency. $\langle\Gamma\rangle$ 
is the average phonon line width, which can be written as
\begin{equation}
\langle\Gamma\rangle = {\sum_{\Bk,\Bk'} \sum_{m,n} \Gamma\ns_{nm}(\Bk,\Bk')\, S\ns_{nm}
(\Bk,\Bk')\over \sum_{\Bk,\Bk'}\sum_{m,n} S\ns_{nm} (\Bk,\Bk') } 
\label{eq:Gamma_ave}
\end{equation}
with
\begin{equation}
\begin{split}
S\ns_{nm} (\Bk,\Bk') &= \delta\big(\mu - E_n(\Bk)\big)\, \delta\big(\mu - E_m(\Bk')\big) \\
\Gamma_{nm}(\Bk,\Bk') &= {\hbar\over 2} \sum_{\Btau,i} m^{-1}_\Btau\, \Tr\!\Big( P_n(\Bk) 
\big[\chi\ns_\Btau\, f\ns_i(\Bk')\\ 
&\hskip-0.3in - f\ns_i(\Bk)\, \chi\ns_\Btau \big] P_{m}(\Bk') 
\big[ \chi\ns_\Btau f\ns_i(\Bk) - f\ns_i(\Bk') \chi\ns_\Btau \big]\Big) \ ,
\label{Main_EQ:Gamma general expr} 
\end{split}
\end{equation}
and where the $(\Bk,\Bk')$ sums span the first Brillouin zone unless otherwise specified. In the expression \cref{Main_EQ:Gamma general expr}, we have the sublattice $\Btau$, $m_{\Btau}$ is the mass of the ion at $\Btau$, with $i \in \{1,\dots,d\}$. From \cref{Main_EQ:Gamma general expr}, we see $\Gamma_{nm}(\Bk,\Bk')$ is a functional of the matrix $f\ns_i(\Bk)$. As a consequence of \cref{Main_EQ:Gaussian Approximation}, we may split $f\ns_i(\Bk)$, as given by \cref{Main_EQ:f_Splitting}. As such, we may split $\Gamma_{n,m}(\Bk,\Bk')$ as
\begin{align}
\Gamma_{nm}^{X_1, X_2}(\Bk,\Bk') &= {\hbar\over 2 m} \sum_{\Btau,i} \Tr\!\bigg( 
P_n(\Bk) \Big[ \chi\ns_\Btau f_i^{X_1}(\Bk') \\
&\hskip -0.5in - f_i^{X_1}(\Bk) \chi\ns_\Btau \Big]\, P_m(\Bk') \Big[
\chi\ns_\Btau f_i^{X_2}(\Bk) - f_i^{X_2}(\Bk') \chi\ns_\Btau \Big] \bigg)\ ,\nonumber
\end{align}
where $X_1, X_2 \in\{E,{\rm geo}\}$.  As such, we find
\begin{equation}
\begin{split}
\la_E &= {2 D(\mu)\over\hbar N}\,{\langle\Gamma\rangle^{E,E}\over\langle\omega^2\rangle}\\
\la_{\rm geo} &= {2 D(\mu)\over\hbar N}\,{\langle\Gamma\rangle^{\rm geo,geo}\over
\langle\omega^2\rangle} \\
\la_{E,{\rm geo}} &={2 D(\mu)\over\hbar N\langle\omega^2\rangle}\,\Big(
\langle\Gamma\rangle^{E,{\rm geo}}+\langle\Gamma\rangle^{{\rm geo},E}\Big) \\
\end{split}
\end{equation}
where $\left\langle \Gamma \right\rangle^{X_1,X_2}$ is defined by replacing
$\Gamma_{nm}(\Bk,\Bk')$ in \cref{eq:Gamma_ave} by $\Gamma_{nm}^{X_1,X_2}(\Bk,\Bk')$.
%
In the following, we will focus on the case where we only have two sublattices in one unit cell and one spinless orbital per atom.
In this case, symmetries can provide significant simplification of the expression.
In particular, two types of symmetries are important for the two-band models of interest: (i) point group symmetries that do not allow invariant representations in the position space, which act on the sublattice as identity, and (ii) the combination of inversion and time-reversal symmetries that switches the two sublattices.
In the presence of both type of symmetries, it is shown that $\lambda_{E,{\rm geo}}=0$, 
\begin{equation}
\lambda_{E} = {\sOmega\over(2\pi)^d}\cdot{\gamma^2\over m \langle \omega^2 \rangle} \sum_n \int\limits_{\textsf{FS}_n}\!\! d\sSigma_{n,\Bk} \> \big|\Bnabla_{\!\Bk} E_n(\Bk) \big|\ , 
\label{Main_EQ:Lambda_E}
\end{equation}
and
\begin{equation}
\lambda_{\rm geo} = {\sOmega\over(2\pi)^d}\cdot{\gamma^2\over m \langle \omega^2 \rangle} \sum_n \int\limits_{\textsf{FS}_n}\!\! d\sSigma_{n,\Bk} \> {\Delta E^2(\Bk)\over|\Bnabla_{\!\Bk} E_n(\Bk)|}
\Tr\!\big[g_n(\Bk)\big]\ ,
\label{Main_EQ:Lambda_Geo}
\end{equation}
with $\Delta E^2(\Bk)=\big| E_+(\Bk) - E_-(\Bk)\big|^2$, where $E_\pm(\Bk)$ is the upper (lower) dispersion in the two-band model. $\textsf{FS}_n$ is the Fermi surface generated by $E_n(\Bk) = \mu$, where $n = \pm$, $\sOmega$ is the volume of the 
unit cell, and $d\sSigma_{n,\Bk}$ is the differential
surface area element on $\textsf{FS}_n$. We also have that $g_n(\Bk)$ is the
quantum metric for band $n$, given by
\begin{align}
[g_n(\Bk)]_{ij} & = \Half\, \Tr\!\!\left[ {\pz P_n(\Bk)\over\pz k_i} \>
{\pz P_n(\Bk)\over\pz k_j}\right] \label{Main_EQ:FSM}\\
& = \Half\, \Tr\!\!\left[ {\pz P_n(\Bk)\over\pz k_i} \> P_n(\Bk) \>
{\pz P_n(\Bk)\over\pz k_j} + (i\leftrightarrow j)\right]\ . \nonumber
\end{align}
(See \cref{Appendix: Introducing EPC} for details.)
In a generic model, we can have other terms in $\la_{X_1,X_2}$, where $X_{1,2}\in\{E,{\rm geo}\}$.
Nevertheless, the special conditions will be satisfied by the 2D Checkerboard model and the hyper-diamond model as discussed in \cref{Pap_SS:CB Symmetries,Pap_SS:HypD Symmetries}.

\section{Geometric Enhancement in the Checkerboard Model \label{Main: Checkerboard Enhancement}}

We first examine the 2D checkerboard model~\cite{sun_time-reversal_2008, sun_topological_2009, li_magic-angle_2022}. The model consists of two spinless orbitals per unit cell, at sublattices $\Btau\ns_1 = -\frac{1}{4} (\Ba\ns_1 + \Ba\ns_2)$ and
$\Btau\ns_2 = \frac{1}{4} (\Ba\ns_1 + \Ba\ns_2)$, where $\Ba_j = a\hh \hat{\bsl{e}}_j$ are the primitive direct lattice vectors. We set the inversion center at the origin between the two 
orbitals, and set the lattice constant $a=1$ for simplicity.
\begin{figure}[H]
\centering
\makebox[-0.5\textwidth]{\textbf{(b)}}%
\makebox[0\textwidth]{\textbf{(a)}}\\
\includegraphics[width=0.5\textwidth]{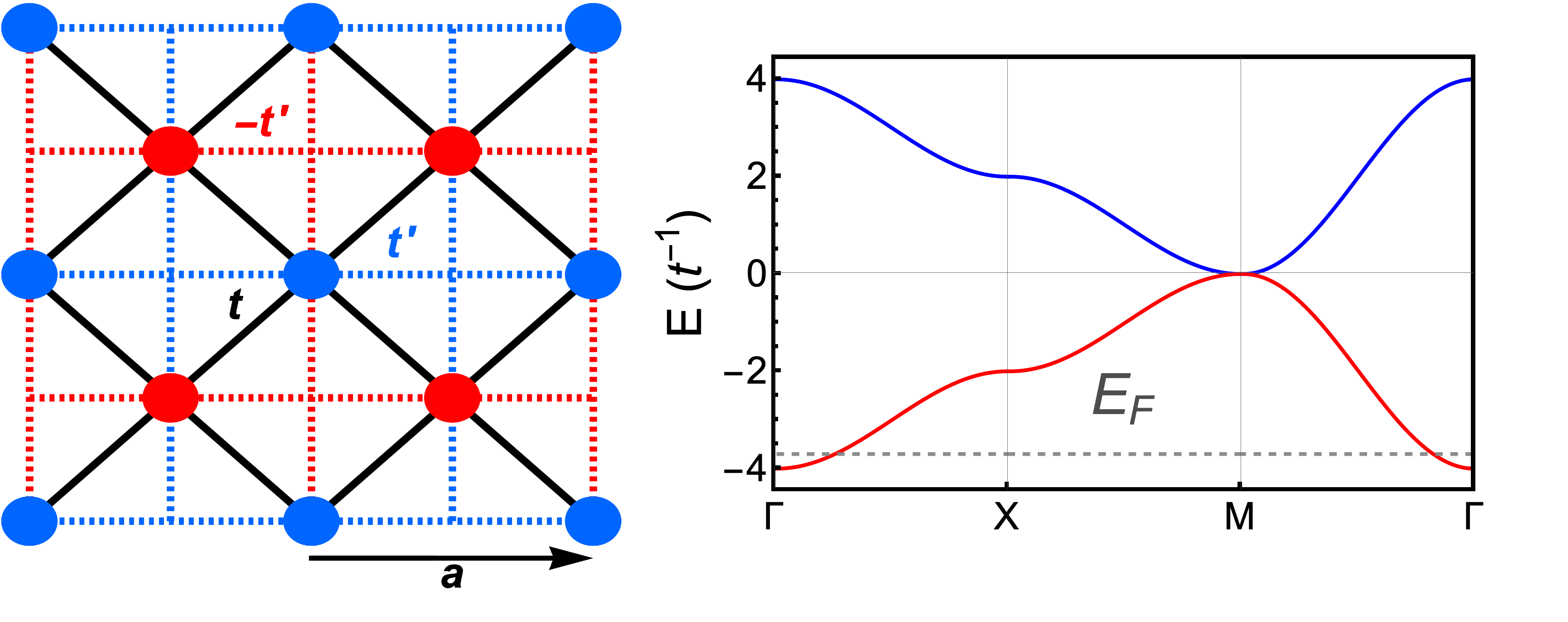}
\caption{(a) Schematic of the checkerboard model. Blue and red sides correspond to sublattices $\Btau\ns_1$ and $\Btau\ns_2$, respectively. These sublattices are connected by nearest-neighbor hopping (black solid lines) and next-nearest-neighbor hopping (blue and red dashed lines). Note that the amplitude of the next-nearest-neighbor hopping is opposite for the two distinct sublattices. The lattice constant is $a$. (b) Band structure of the checkerboard model along the high-symmetry line $\Gamma - X - M - \Gamma$. The dispersion is given in units of the nearest-neighbor hopping coefficient $t$. The Fermi level (gray dashed line) lies in the lower band, and corresponds to the $E_{\rm F} < 0$ case.}
\label{Main_Fig:CB_plots}
\end{figure}

\begin{figure}[H]
\centering
\includegraphics[width=0.5\textwidth]{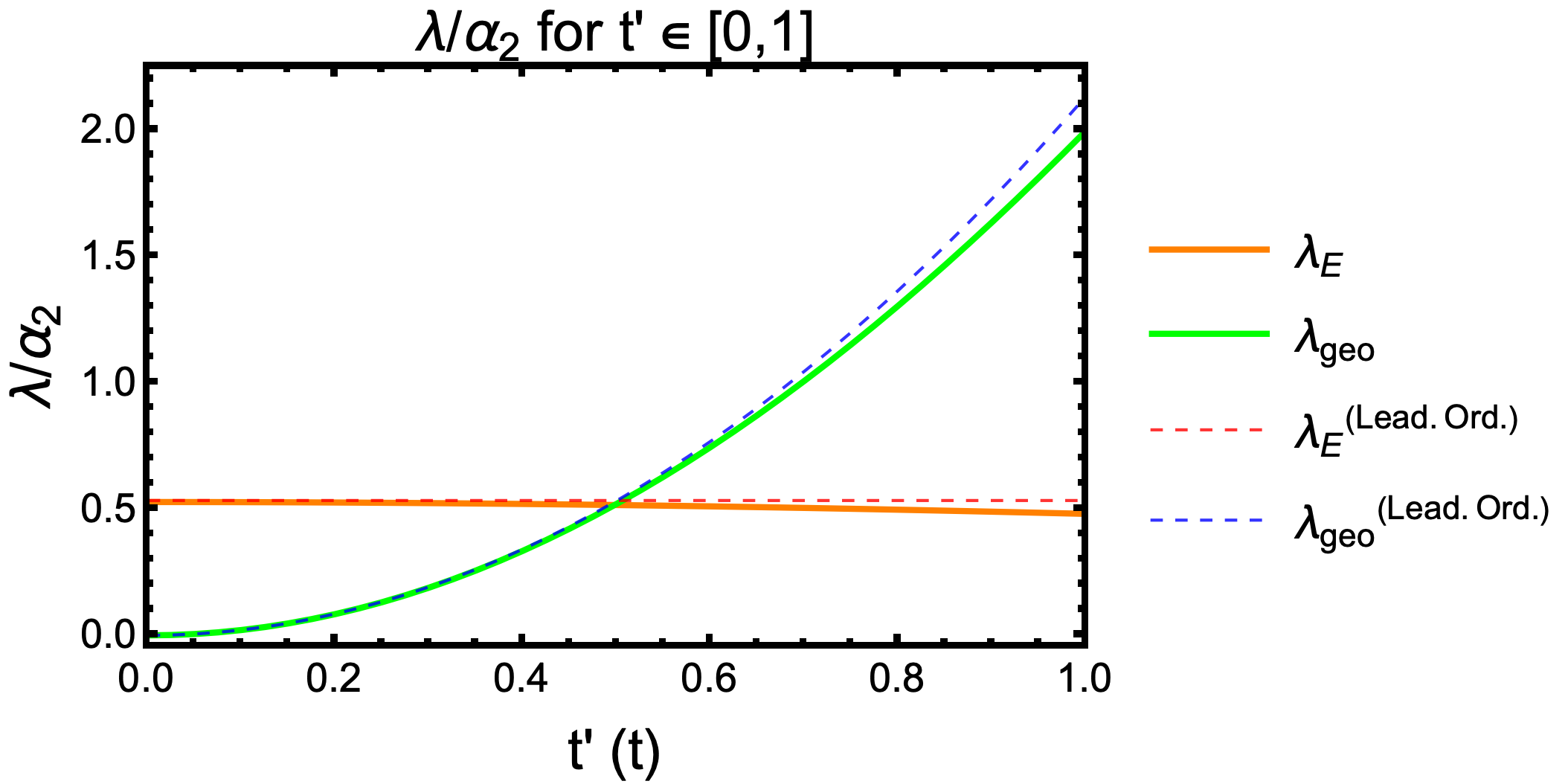}
\caption{Numerical calculations of $\la_E$ and $\la_{\rm geo}$, given by \cref{Main_EQ:Lambda_E,Main_EQ:Lambda_Geo}, and their leading-order approximations according to \cref{Main_EQ:LeadOrd_lambdas}, in units of $\al_2$, as a function of $t′\in \left[0,1\right]$ in units of the nearest-neighbor hopping t. Solid lines represent the exact numerical calculations, with dashed lines representing their leading-order approximations.}
\label{Main_Fig:Geometric Enhancement}
\end{figure}

The checkerboard model has a two-band Bloch Hamiltonian $h(\Bk) = \Bd(\Bk) \cdot \mbf{\sigma}$, 
where $\mbf{\sigma}=(\sigma^x,\sigma^y,\sigma^z)$ are the Pauli matrices and
$\Bd(\Bk) = d_x(\Bk)\,\hat{\bsl{e}}_x + d_z(\Bk)\,\hat{\bsl{e}}_z$, with  
$d_x(\Bk) = 4 t \cos(\half k_x) \cos (\half k_y)$ and $d_z(\Bk) = -2 t'\big[ \cos(k_x) - \cos(k_y)
\big]$. The checkerboard model is symmetric under the point group symmetry $D_{4h}$. 
Additionally, it is invariant under the operations of parity $\P$ and time-reversal $\T$. We may 
show that the symmetry properties of the checkerboard model require that the EPC contributions 
have the form as given by \cref{Main_EQ:Lambda_E,Main_EQ:Lambda_Geo}, as addressed in 
\cref{Pap_SS:CB Symmetries}.
The dispersion has the form $E_\pm(\Bk) = \pm | \Bd(\Bk) |$, and the projectors take the form $P_{\pm}(\Bk) = \frac{1}{2} \sigma^0 \pm \frac{1}{2} \hat{\Bd}(\Bk) \cdot \mbf{\sigma}$, where 
$\sigma^0$ is identity and $\hat{\Bd}(\Bk) \equiv \Bd(\Bk)/\big|\Bd(\Bk)\big|$. As a result, 
we can derive a general expression for the diagonal elements of the quantum metric in a 
two-band model, {\it viz.\/}
\begin{align}
[g_\pm(\Bk)]_{ii} & = {1\over 4} \big| \pz\ns_{k_i} \hat{\Bd}(\Bk) \big|^2 
\label{Main:FSM in terms of d} \\
&\hskip-0.2in = {1\over 4 \big| \Bd(\Bk) \big|^2} \bigg|\pz\ns_{k_i} \Bd(\Bk) -
\hat{\Bd}(\Bk) \left( \hat{\Bd}(\Bk) \cdot \pz\ns_{k_i}\Bd(\Bk) \right) \bigg|^2\ . \nonumber
\end{align}
We assume the hopping parameters $t$ and $t'$ are non-negative and the chemical potential $\mu < 0$.
In the vicinity of the $\Gamma$ point, where $\mu$ is close to $-4 t$, we find that the dispersion
$E_\pm(\Bk) = \pm 4t\hh \big( 1 - \frac{1}{8}|\Bk|^2 \big) + O(k^4)$. As such, $\la_E$ is independent of $t'$ to the leading order. However, since $\Tr\!\big[ g_\pm(\Bk)\big] = {t'}^2\Bk^2/{d_x^2(\Bk)}$ to lowest order, $\la_{\rm geo}$ tunes with $t'$.  To leading order we obtain
\eqa{
\lambda_E=t\CI\ , \qquad \lambda_{\rm geo}=4{t'}^2\CI/t\ ,
\label{Main_EQ:LeadOrd_lambdas}
}
where
\begin{equation}
\CI= \al_2\! \int\limits_{\textsf{FS}_-}\!\! d\sSigma_{-,\Bk} \>|\Bk| \quad,\quad
\alpha\ns_d={\sOmega\over(2\pi)^d}\cdot{\gamma^2\over m \langle \omega^2 \rangle}\ .
\end{equation}
Thus we have realized a model where we can enhance the strength of EPC by enhancing the geometric properties while keeping the energetic properties unchanged to the leading order. We verify these results in \cref{Main_Fig:Geometric Enhancement}, where we plot both $\la_E$ and $\la_{\rm geo}$ to the leading order, as well as their exact expression at $\mu = -3.95 t$. On average, the leading order approximations are accurate to $2.37\%$ and $3.37\%$ for $\la_E$ and $\la_{\rm geo}$, respectively. 

In our model, the geometric contribution to strength of EPC can be tuned with the 
next-nearest-neighbor hopping $t'$ in the region of the $\Gamma$ point, while keeping 
the energetic contribution nearly constant. For $t' > t/2$, the geometric contribution 
is larger than the energetic contribution. 

\section{Geometric Contribution to $\la$ in $d$ Dimensional hyper-diamond \label{Main: Geo in d Dim Hyperdiam}}

We now examine the geometric contribution in the generalized $d$ dimensional hyper-diamond model \cite{rachel_strain-induced_2016, kohler_nodal_2024}. In $d$ dimensions, each sublattice has $d+1$ nearest-neighbors. We may iteratively generate these nearest-neighbor vectors, which we denote as $\hat{\mbf{\delta}}^{(d)}_l$, where $l \in \{1,\dots,d+1\}$. We begin in 1D by defining $\hat{\mbf{\delta}}_1^{(1)} = \bsl{\hat{e}}_1$ and $\hat{\mbf{\delta}}_2^{(1)} = -\bsl{\hat{e}}_1$. In dimensions $d>1$, we define the vectors $\hat{\mbf{\delta}}^{(d)}_l$ as
\eqa{
\hat{\mbf{\delta}}^{(d)}_l = \begin{cases}
    \sqrt{1-d^{-2}}\,\hat{\mbf{\delta}}^{(d-1)}_l - d^{-1} \hat{\bsl{e}}_d & \text{if } 
    1\le l \le d\\
    \hat{\bsl{e}}_d & \text{if } l = d+1\ . \\
\end{cases}
\label{Pap_EQ:NN hyper-diamond Algorithm}
}
We can show that $\hat{\mbf{\delta}}_l^{(d)} \cdot \hat{\mbf{\delta}}_{l'}^{(d)} = -d^{-1} + (1+d^{-1}) \delta_{ll'}$ and $\sum_{l=1}^{d+1} \hat{\mbf{\delta}}_l^{(d)} = 0$. We can use $\hat{\mbf{\delta}}_l^{(d)}$ to construct the primitive lattice vectors $\Ba_j^{(d)} = \hat{\mbf{\delta}}_{d+1}^{(d)} - \hat{\mbf{\delta}}_j^{(d)} = \hat{\bsl{e}}_d - \hat{\mbf{\delta}}_j^{(d)}$ for $j \in \{1, \dots, d\}$. We can then construct the primitive reciprocal lattice vectors $\Bb_i^{(d)}$ using $\Ba_j^{(d)} \cdot \Bb_i^{(d)} = 2\pi\hh\delta_{ji}$, leading to  $\Bb_i^{(d)} \cdot \hat{\mbf{\delta}}_l^{(d)} = 2\pi(d+1)^{-1} - 2\pi\hh\delta_{li}\equiv 2\pi \beta\ns_{li}$, a $(d+1)\times d$ matrix. We assume two spinless orbitals per unit cell with 
atomic positions
\eqa{
\Btau_{\mu}^{(d)} = {(-1)^\mu\over 2 (d+1)} \sum_{j=1}^d \Ba_j^{(d)} = \Half (-1)^\mu\,
\hat{\bsl{e}}_d\ .
}
for $\mu = 1,2$, with the inversion center set at the origin between the two orbitals. With the nearest neighboring hopping $t$, the $d$ dimensional hyper-diamond tight-binding model has the
momentum space form $h(\Bk)=t\textsf{Re}\gamma(\Bk)\,\sigma^x -t\textsf{Im}\gamma(\Bk)\,\sigma^y$,
where $\gamma(\Bk) = \sum_{l=1}^{d+1} e^{-\ii \Bk \cdot \hat{\mbf{\delta}}^{(d)}_l}$.
Without loss of generality, we set the hopping $t$ to be 1 eV.
 This Bloch Hamiltonian can be re-expressed in the two-band form $h(\Bk) = \Bd(\Bk) \cdot \mbf{\sigma}$, where $\Bd(\Bk) = d_x(\Bk)\,\hat{\bsl{e}}_x + d_y(\Bk)\,\hat{\bsl{e}}_y$, with 
$d_x(\Bk) = \sum_{l=1}^{d+1}\cos\!\Big(\sum_{i=1}^d \beta\ns_{li}\, k\ns_i\Big)$ and $d_y(\Bk) = \sum_{l=1}^{d+1}\sin\!\Big(\sum_{i=1}^d \beta\ns_{li}\, k\ns_i\Big)$, where
$\Bk =(2\pi)^{-1}\sum_{i=1}^d k\ns_i\hh\Bb_i^{(d)}$ and $k_i \in [0,2\pi)$. 
This two-band Hamiltonian is invariant under $\PT$ symmetry. 

We may describe the symmetries of the $d$ dimensional lattice by defining a $C_3$ symmetry plane given by $\Pi_i = \textsf{span} \big\{ \mbf{\al}_i^{(d)}, \mbf{\al}_{i+1}^{(d)} \big\}$, where $\mbf{\al}_i^{(d)} = \hat{\mbf{\delta}}_i^{(d)} - \hat{\mbf{\delta}}_{i + 1}^{(d)}$ for $i \in \{1,\dots,d-1\}$. One can derive the $d \times d$ matrix representation of these rotations $C_3^{\Pi_i}$ by orthonormalizing $\mbf{\al}_i^{(d)}$ and $\mbf{\al}_{i+1}^{(d)}$, and prove that they do preserve the hyper-diamond symmetries by acting on the basis $\BR^{(d)} + \Btau_\mu^{(d)}$ (see details in \cref{Pap_SS:HypD Symmetries}). We generate $d-1$ symmetry operations using this process, all of which act on the sublattice as identity.
As the origin is set between the two orbitals where $\Btau_1^{(d)} = -\Btau_2^{(d)}$, we have that $\PT$ switches the sublattices. As a result, the EPC constant $\la$ is constrained to the form of \cref{Main_EQ:Lambda_E,Main_EQ:Lambda_Geo}. The details of these calculations are provided in \cref{Pap_SS:HypD Symmetries}.

\begin{figure*}[t]
\centering
\begin{subfigure}{0.32\textwidth} 
\includegraphics[width=\textwidth]{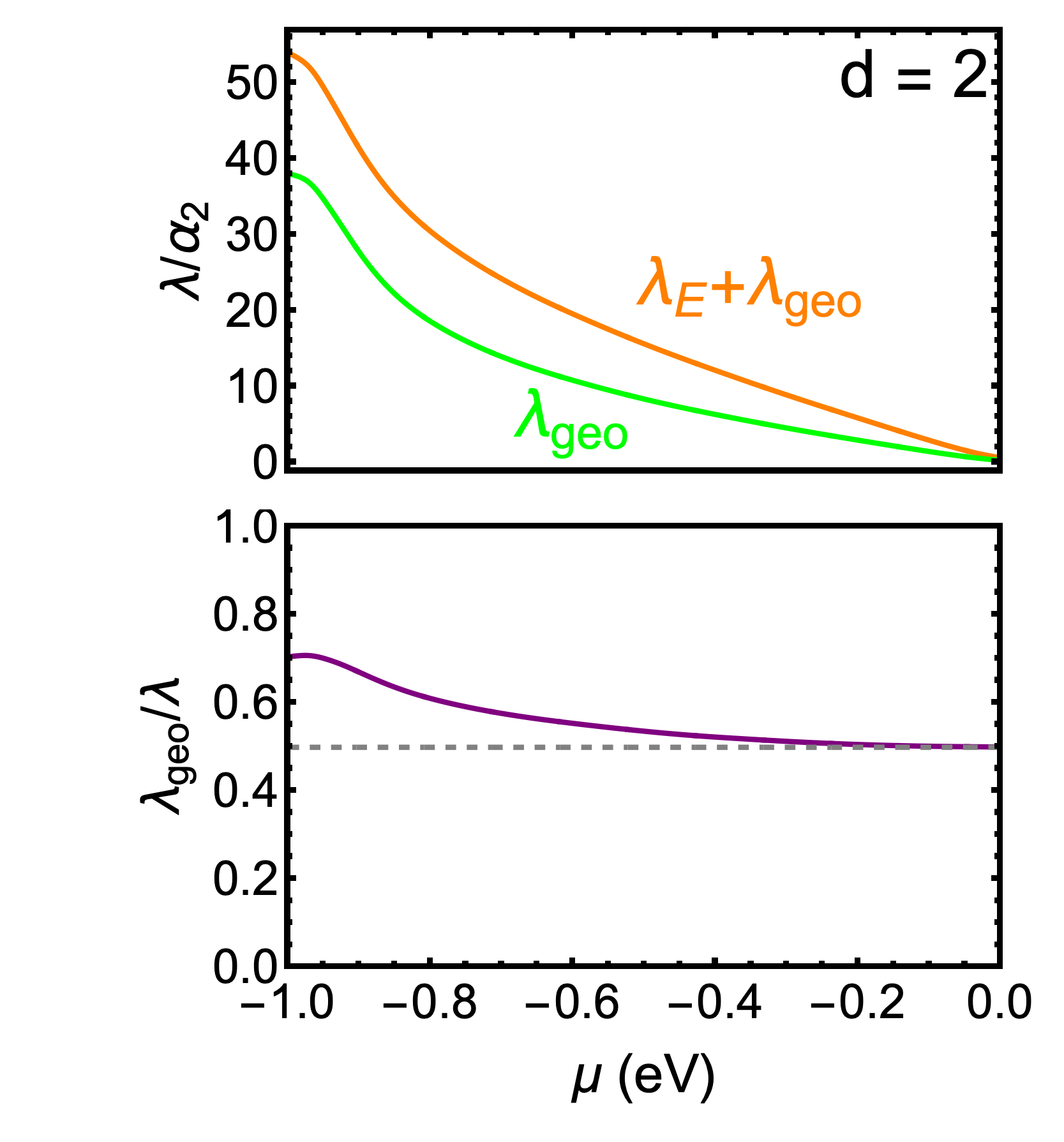}
\caption{}
\label{App_Subfig: HypD 2d}
\end{subfigure}
\hfill
\begin{subfigure}{0.32\textwidth} 
\includegraphics[width=\textwidth]{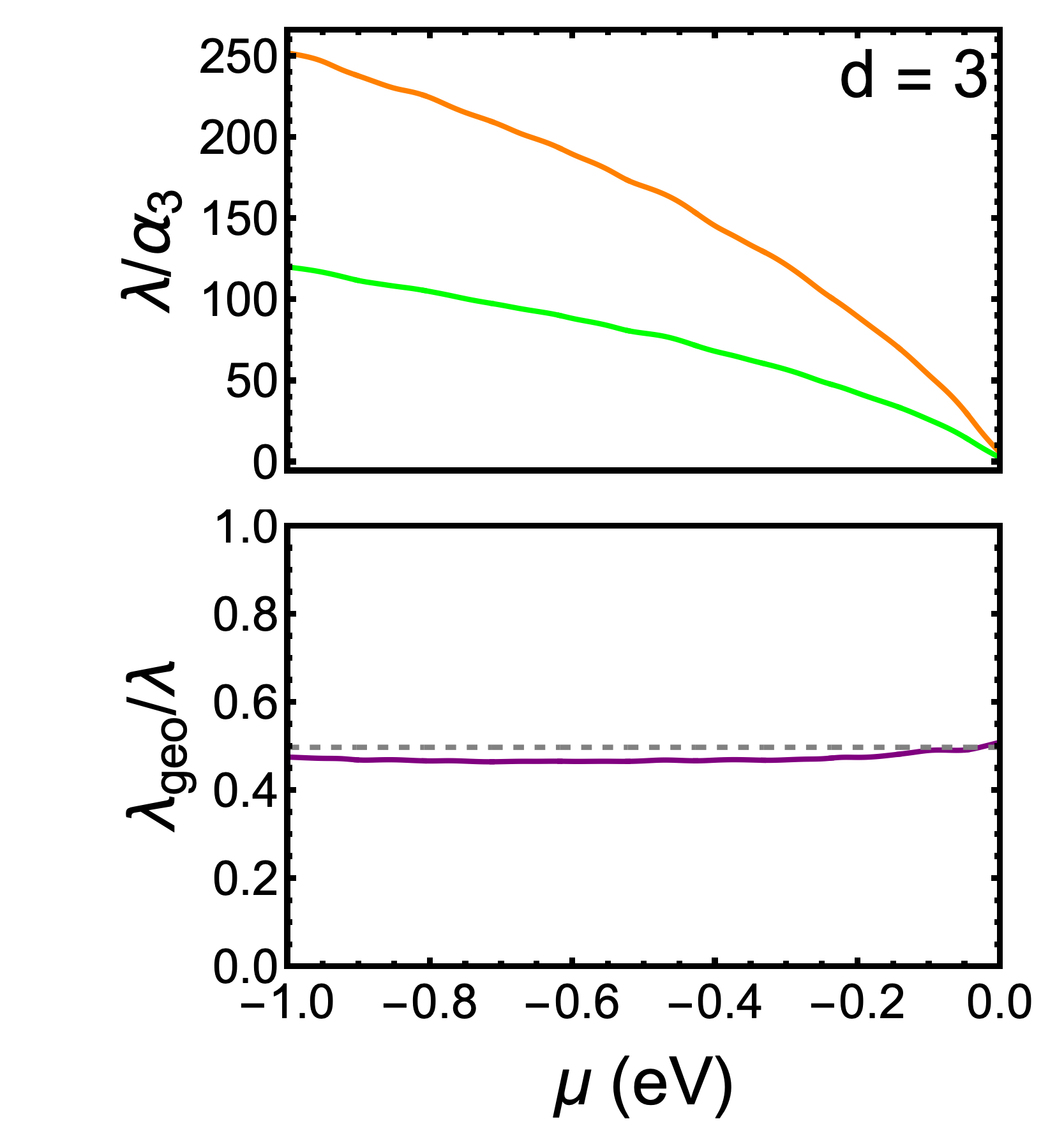}
\caption{}
\label{App_Subfig: HypD 3d}
\end{subfigure}
\hfill
\begin{subfigure}{0.32\textwidth} 
\includegraphics[width=\textwidth]{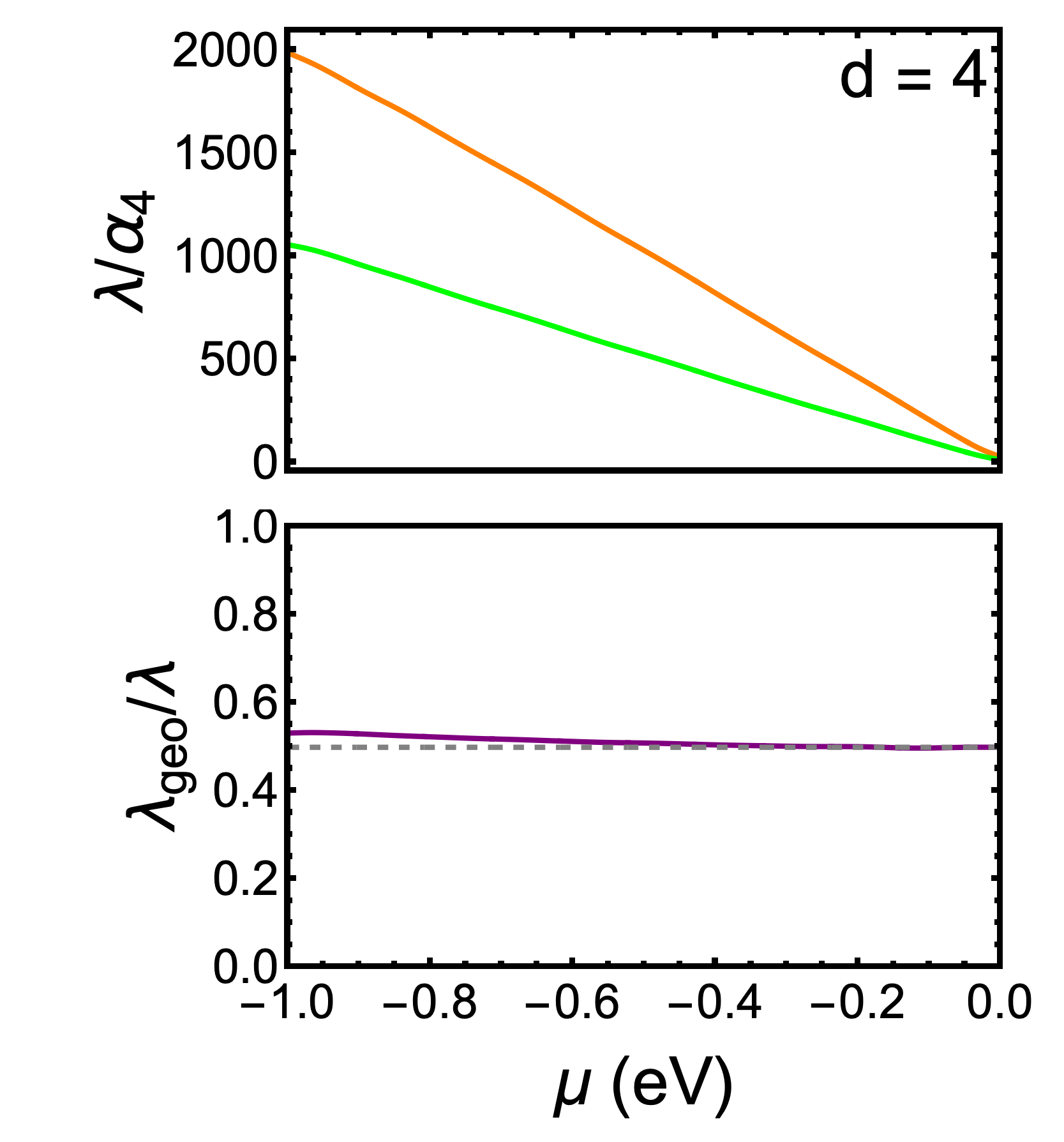}
\caption{}
\label{App_Subfig: HypD 4d}
\end{subfigure}
\vskip 1em
\begin{center} 
\begin{subfigure}{0.32\textwidth} 
\includegraphics[width=\textwidth]{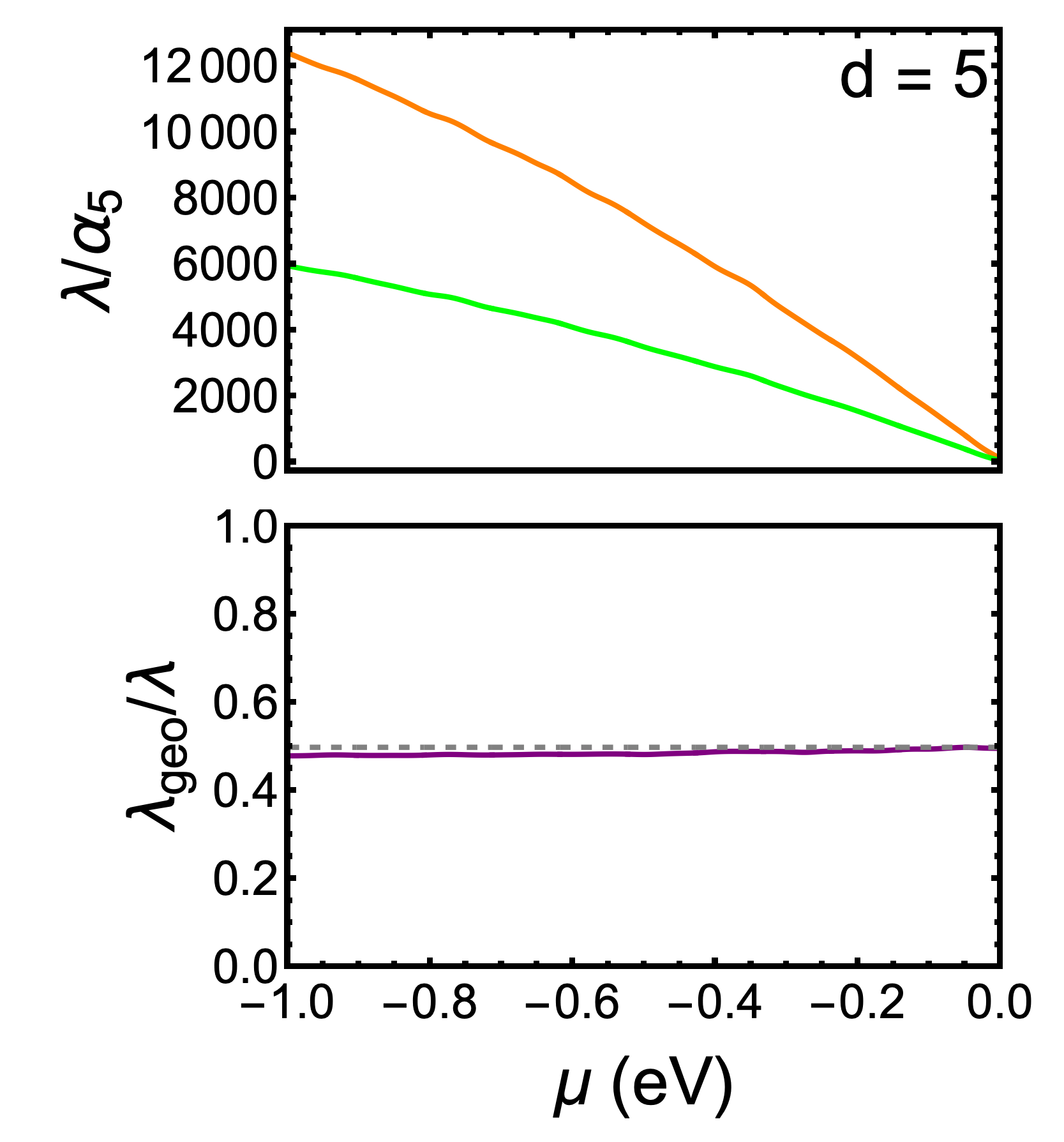}
\caption{}
\label{App_Subfig: HypD 5d}
\end{subfigure}
\begin{subfigure}{0.32\textwidth} 
\includegraphics[width=\textwidth]{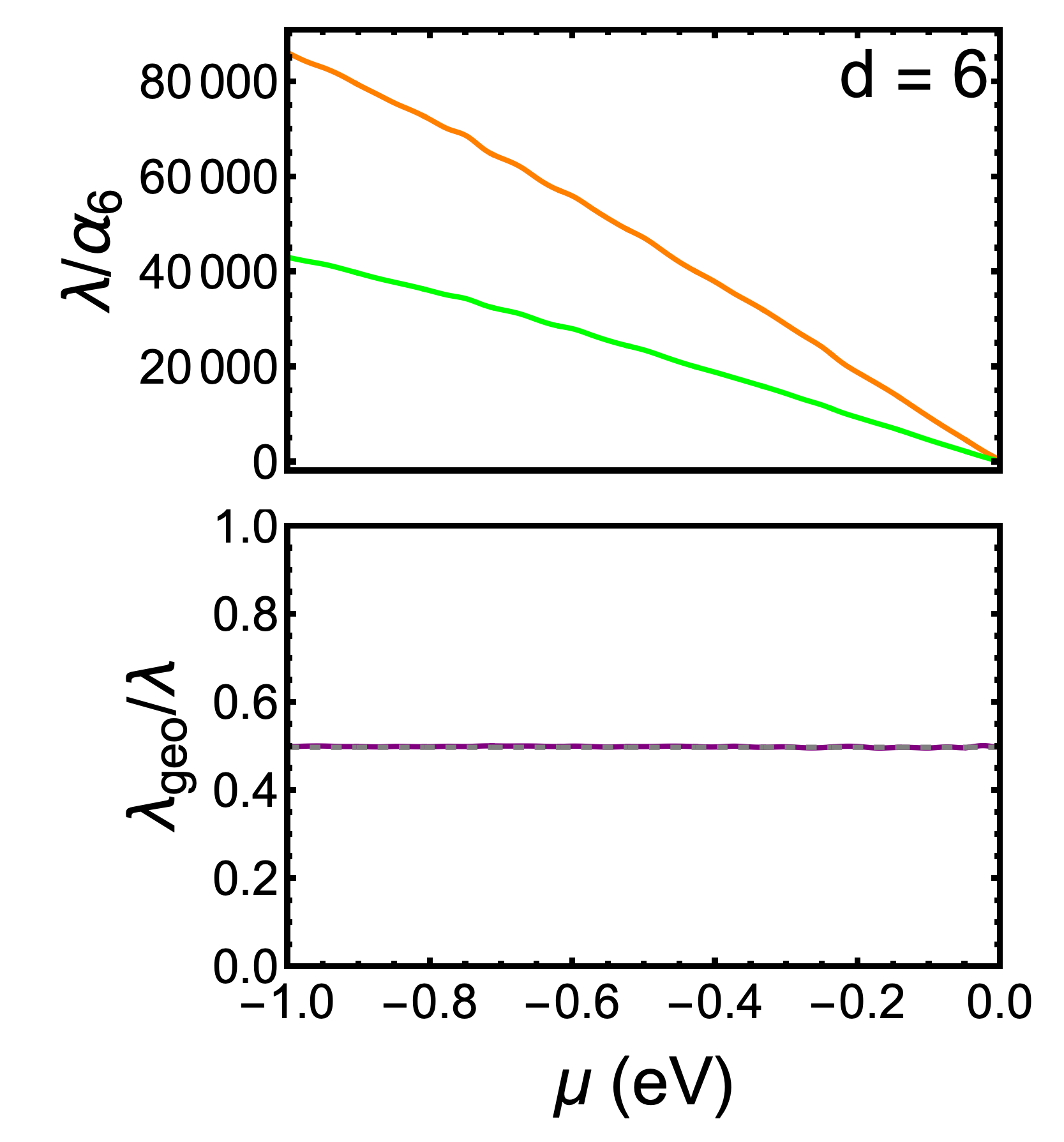}
\caption{}
\label{App_Subfig: HypD 6d}
\end{subfigure}
\end{center} 
\caption{Numerical calculations of $\la = \la_E + \la_{\rm geo}$ and $\la_{\rm geo}$, in units of $\al_d$, and their ratio $\la_{\rm geo}/\la$, as a function of chemical potential $\mu\in[−1~\eV,0~\eV]$, for the hyper-diamond model in dimensions $d=2$ through $d=6$. In each panel, the upper plot shows $\la_E + \la_{\rm geo}$ (orange) and $\la_{\rm geo}$ (green), and the lower plot shows $\la_{\rm geo}/\la$ (purple). The gray dashed line denotes $\la_{\rm geo}/\la = 1/2$, the analytically derived limit as $\mu$ approaches 0. (a) $d=2$. (b) $d=3$. (c) $d=4$. (d) $d=5$. (e) $d=6$.}
\label{Main_Fig:hypd_Plots}
\end{figure*}

We now consider the ratio $\la_{\rm geo}/(\la_E + \la_{\rm geo})$ as $\mu \rightarrow 0$ in 
$d \geq 2$ dimensions. We write $\Bk = \sum_{i=1}^d k_i\hh \Bb_i/2\pi$, where each
$k_i \in \left[ 0, 2\pi \right)$. We then expand $\gamma(\Bk)$ about 
$\Bk \equiv \Bp + \Bq$ for $|\Bq| \ll 1$. We find that $\la_E \rightarrow \la_{\rm geo} 
\rightarrow \int\limits_{\textsf{FS}_0} \!\! dV_\Bp \frac{\pi \lp \bsl{c}_{\Bp}^2 + 
\bsl{s}_{\Bp}^2\rp}{\sqrt{|\bsl{c}_{\Bp}|^2 |\bsl{s}_{\Bp}|^2 - |\bsl{c}_{\Bp} \cdot \bsl{s}_{\Bp}|^2}}$ as $\mu\rightarrow 0$, where $dV_\Bp$ is the $d-2$ volume element at 
$\Bp$, $\textsf{FS}_0$ is the Fermi surface for $\mu = 0$, $\bsl{c}_{\Bp} = \sum_{i=1}^d \cos(p_i)\, \Ba_i$,
and $\bsl{s}_{\Bp} = \sum_{i=1}^d \sin(p_i)\, \Ba_i$ (omitting the $(d)$ superscript in this notation). As such, the ratio $\la_{\rm geo}/(\la_E + \la_{\rm geo})$ in $d$ dimensions approaches 
$1/2$ as $\mu\to 0$. The details of this proof are shown in 
\cref{Sub Appendix: Proof of Ratio in d Dim}. To verify our results, we numerically calculate 
the ratio $\la_{\rm geo}/(\la_E + \la_{\rm geo})$ for $d = 2,\dots,6$, the results of which can 
be seen in \cref{Main_Fig:hypd_Plots}. We evaluate $\la$ in the range $\mu \in [-1~\eV,0~\eV]$ \cite{castro_neto_electronic_2009}.
We find that in dimensions $d\in\{2,\dots,6\}$ that the ratio  $\la_{\rm geo}/(\la_E + \la_{\rm geo})$ approaches 1/2 as $\mu\to 0$, and both contributions individually approach zero in the same limit.
Interestingly, the deviation from $1/2$ when $\mu$ deviates from zero is suppressed as the dimension increases. 
These results generalize the findings of \refcite{yu_nontrivial_2024}, confirming the ratio of $1/2$ beyond the 2D case into any dimension.

\section{Discussion and Conclusion}

In this work, we have systematically investigated the role of quantum geometry in the electron-phonon coupling (EPC) constant $\lambda$ using the Gaussian approximation \cite{yu_nontrivial_2024} in both the checkerboard~\cite{sun_time-reversal_2008, sun_topological_2009, li_magic-angle_2022} and hyper-diamond~\cite{rachel_strain-induced_2016, kohler_nodal_2024} models. Our analysis of the 2D checkerboard model reveals that the geometric contribution to $\lambda$ is highly tunable via the next-nearest-neighbor hopping parameter $t'$. Crucially, we found that in the vicinity of the $\Gamma$ point, the geometric contribution can be enhanced while keeping the energetic contribution nearly constant by increasing $t'$. This result provides a concrete proof-of-principle that EPC can be strengthened through geometric engineering, potentially leading to higher superconducting critical temperatures.

Furthermore, we generalized the result for graphene in \refcite{yu_nontrivial_2024} to arbitrary
dimensions using the hyper-diamond model. We analytically proved and numerically verified that as
the system approaches charge neutrality ($\mu\to 0$), the geometric contribution to the EPC
constant approaches exactly half of the total interaction strength, regardless of the spatial
dimension $d$.

\section{Acknowledgments}

We thank Aris Alexandradinata, Jonah Herzog-Arbeitman, and B. Andrei Bernevig for helpful discussions. This work is supported by startup funds at University of Florida. T.C.'s work is also supported by the University of Florida Graduate School's Grinter Fellowship, as well as the University of Florida Center for Condensed Matter sciences (CCMS) Graduate Fellowship.

\clearpage

\bibliography{hyperdiamond}

\appendix
\onecolumngrid

\tableofcontents

\section{Reviewing Electron-Phonon Coupling and the Gaussian Approximation \label{Appendix: Introducing EPC}}

In this section, we review the concept of the electron-phonon coupling (EPC), following \refcite{yu_nontrivial_2024}. We will only consider the spinless case and will consider one orbital per atom, which is the case for both the checkerboard model and hyper-diamond model, as discussed in \cref{Appendix: Geo Enhancement in CB,Appendix: hyper-diamond Lattice}, respectively.
Then, our tight-binding Hamiltonian is given by
\eqa{
H_{\mathrm{el}} = \sum_{\BR, \BR'} \sum_{\Btau, \Btau'} t_{\Btau \Btau'}(\BR+\Btau- \BR'-\Btau') c_{\BR+\Btau}^\dagger c\nd_{\BR'+\Btau'}\ ,
\label{Pap_EQ:Tight-Binding Hamiltonian}
}
where $\BR$ is a lattice vector, $\Btau$ a sublattice, $c_{\BR+\Btau}^\dagger$ creates an electron in the spinless orbital at $\BR+\Btau$, and $t_{\Btau \Btau'}(\BR+\Btau - \BR'-\Btau')$ is our hopping function. The general Hamiltonian, which takes into account atomic motion under the two-center approximation, reads
\eqa{
H_{\mathrm{full}} = \sum_{\BR, \BR'} \sum_{\Btau, \Btau'}  t_{\Btau \Btau'}(\BDelta_{\BR \BR'}^{\Btau \Btau'}) c_{\BR+\Btau}^\dagger c\nd_{\BR'+\Btau'}\ ,
\label{Pap_EQ:Most gen Hamiltonian form}
}
with
\eqa{
\BDelta_{\BR \BR'}^{\Btau \Btau'} = \BR + \Btau + \Bu_{\BR+\Btau} - \BR' - \Btau' - \Bu_{\BR'+\Btau'}\ ,
}
where $\Bu_{\BR+\Btau}$ is the ionic motion at $\BR+\Btau$. The tight-binding model \cref{Pap_EQ:Tight-Binding Hamiltonian} is the zeroth-order term in the expansion of \cref{Pap_EQ:Most gen Hamiltonian form} with respect to $\Bu_{\BR+\Btau} - \Bu_{\BR'+\Btau'}$. The first-order term is the leading-order contribution to the electron-phonon coupling (EPC), which has the form
\begin{align}
& H_{\mathrm{el-ph}} = \sum_{\BR,\Btau,\BR',\Btau'} (\Bu_{\BR+\Btau} - \Bu_{\BR'+\Btau'}) \cdot \BQ_{\BR \BR'}^{\Btau \Btau'}\, c_{\BR+\Btau}^\dagger c_{\BR'+\Btau'}\nd\ , \\
& \BQ_{\BR \BR'}^{\Btau \Btau'} = \frac{\del t_{\Btau \Btau'}(\Br)}{\del \Br} \bigg|_{\Br = \BR + \Btau - \BR' - \Btau'}\ . \\
\end{align}
This is what we focus on in this work.

\subsection{Gaussian Approximation}

We work under the Gaussian approximation, which assumes the hopping function satisfies
\eqa{
t_{\Btau \Btau'}(\Br) = t_{\Btau \Btau'}^{0} \exp \lp \frac{1}{2} \gamma_{\Btau\Btau'} \Br^2 \rp = t_{\Btau \Btau'}^{0} \exp \lp \frac{1}{2} \gamma \Br^2 \rp\ ,
\label{Pap_EQ:Gaussian Approximation}
} 
where $t_{\Btau \Btau'}^{0}$ and $\gamma_{\Btau\Btau'}$ constants. We choose $\gamma_{\Btau\Btau'} = \gamma$ for simplicity, where $\gamma < 0$ to ensure the decay of the hopping function as $\BR$ increases. We may convert $H_{\mathrm{el-ph}}$ to reciprocal space using the Fourier transformation rule of the basis: 
\eqa{
c_{\BR+\Btau}^\dagger = \frac{1}{\sqrt{N}} \sum_\Bk e^{-\ii \Bk \cdot (\BR + \Btau)} c_{\Bk,\Btau}^\dagger \quad u_{\BR+\Btau, i} = \frac{1}{\sqrt{N}} \sum_{\Bq} e^{\ii \Bq \cdot (\BR + \Btau)} u_{\Bq, \Btau, i}\ ,
\label{Pap_EQ:FT Rule}
}
where $N$ is the number of unit cells.
We implement the transformation, reading
\eqa{
H_{\mathrm{el-ph}}&  = \frac{1}{N^{3/2}}\sum_{\BR,\BR'} \sum_{\Bq} \sum_{\Bk,\Bk'} \sum_{\Btau,\Btau'} \sum_i e^{\ii \Bq \cdot (\BR+\Btau)} u_{\Bq, \Btau, i} \left[\del_{r_i} t_{\Btau \Btau'}(\Br)\right]_{\Br = \BR+\Btau-\BR'-\Btau'} e^{-\ii \Bk \cdot (\BR+\Btau)} c_{\Bk,\Btau}^\dagger e^{\ii \Bk' \cdot (\BR'+\Btau')} c_{\Bk',\Btau'}\nd \\
& \quad  -\frac{1}{N^{3/2}}\sum_{\BR,\BR'} \sum_{\Bq} \sum_{\Bk,\Bk'} \sum_{\Btau,\Btau'} \sum_i e^{\ii \Bq \cdot (\BR'+\Btau')} u_{\Bq, \Btau',i} \left[\del_{r_i} t_{\Btau \Btau'}(\Br)\right]_{\Br = \BR+\Btau-\BR'-\Btau'} e^{-\ii \Bk \cdot (\BR+\Btau)} c_{\Bk,\Btau}^\dagger e^{\ii \Bk' \cdot (\BR'+\Btau')} c_{\Bk',\Btau'}\nd\ , \\
}
with $\del_{r_i} \equiv \frac{\del}{\del r_i}$. We redefine
$\Bk \equiv \Bk_1, \Bk' \equiv \Bk_2, \Btau \equiv \Btau_1, \Btau' \equiv \Btau_2$,
and use $u_{-\Bq, \Btau, i} = u^\dagger_{\Bq, \Btau, i}$ to find
\eqa{
H_{\mathrm{el-ph}} = \frac{1}{\sqrt{N}} \sum_{\Bk_1,\Bk_2} \sum_{\Btau_1,\Btau_2} \sum_i c_{\Bk_1,\Btau_1}^\dagger c_{\Bk_2,\Btau_2}\nd \left[ u_{\Bk_2 - \Bk_1,\Btau_1, i}^\dagger [f_i(\Bk_2)]_{\Btau_1 \Btau_2} - u_{\Bk_2 - \Bk_1,\Btau_2, i}^\dagger [f_i(\Bk_1)]_{\Btau_1 \Btau_2} \right]\ ,
}
where
\eqa{
[f_i(\Bk)]_{\Btau_1 \Btau_2} = \sum_\BR e^{-\ii \Bk \cdot (\BR+\Btau_1 - \Btau_2)} \left[\del_{r_i} t_{\Btau_1 \Btau_2}(\Br)\right]_{\Br = \BR+\Btau_1-\Btau_2}\ .
}
By defining $[\chi_{\Btau}]_{\Btau_1 \Btau_2} = \delta_{\Btau,\Btau_1} \delta_{\Btau_1,\Btau_2}$ the projection matrix onto the sublattice $\Btau$, we may simplify $H_{\mathrm{el-ph}}$:
\eqa{
H_{\mathrm{el-ph}} & = \frac{1}{\sqrt{N}} \sum_{\Bk_1, \Bk_2} \sum_{\Btau,\Btau_1,\Btau_2} \sum_i c_{\Bk_1, \Btau_1}^\dagger c_{\mbf{k}_2, \Btau_2}\nd u_{\Bk_2 - \Bk_1,\Btau, i}^\dagger [ \delta_{\Btau_1,\Btau} [f_i(\Bk_2)]_{\Btau_1 \Btau_2} - [f_i(\Bk_1)]_{\Btau_1 \Btau_2} \delta_{\Btau_2,\Btau}] \\
& = \frac{1}{\sqrt{N}} \sum_{\Bk_1, \Bk_2} \sum_{\Btau,\Btau_1,\Btau_2} \sum_i c_{\Bk_1, \Btau_1}^\dagger [\chi_{\Btau} f_i(\Bk_1) - f_i(\Bk_2) \chi_{\Btau}]_{\Btau_1 \Btau_2} c_{\mbf{k}_2, \Btau_2}\nd u_{\Bk_2 - \Bk_1,\Btau, i}^\dagger \\
& = \frac{1}{\sqrt{N}} \sum_{\Bk_1, \Bk_2} \sum_{\Btau} \sum_i c_{\Bk_1}^\dagger [\chi_{\Btau} f_i(\Bk_1) - f_i(\Bk_2) \chi_{\Btau}] c_{\Bk_2}\nd u_{\Bk_2 - \Bk_1, \Btau, i}^\dagger\ ,
}
where $c_\Bk^\dagger \equiv \lp \dots, c_{\Bk, \Btau}^\dagger, \dots \rp$ is a row vector of creation operators. Under the Gaussian approximation, $\Bnabla_\Br t_{\Btau \Btau'}(\Br) = \gamma \Br t_{\Btau \Btau'}(\Br)$\ . Then
\eqa{
[f_i(\Bk)]_{\Btau_1 \Btau_2} & = \gamma \sum_{\BR} e^{-\ii \Bk \cdot (\BR+\Btau_1 - \Btau_2)} (\BR+\Btau_1 - \Btau_2)_i\, t_{\Btau_1 \Btau_2}(\BR+\Btau_1 - \Btau_2) \\
& = \ii \gamma \del_{k_i} [h(\Bk)]_{\Btau_1 \Btau_2}\ ,
\label{Pap_EQ:f_i in terms of h(k)}
}
where $[h(\Bk)]_{\Btau_1 \Btau_2} = \sum_{\BR} e^{-\ii \Bk \cdot (\BR+\Btau_1 - \Btau_2)} t_{\Btau_1 \Btau_2}(\BR+\Btau_1 - \Btau_2)$ is the electron Hamiltonian and $\del_{k_i} \equiv \frac{\del}{\del k_i}$. Then \cref{Pap_EQ:f_i in terms of h(k)} leads to
\eq{
f_i(\Bk) = \ii \gamma \del_{k_i} h(\Bk)\ .
}
We may write $h(\Bk) = \sum_n E_n(\Bk) P_n(\Bk)$, where $E_n(\Bk)$ is the dispersion of the $n$th band, and $P_n(\Bk)$ is the projector operator, defined as $P_n(\Bk) = U_n(\Bk) U_n\yd(\Bk)$. $U_n(\Bk)$ satisfies $h(\Bk) U_n(\Bk) = E_n(\Bk) U_n(\Bk)$. As such, we can split 
$f\ns_i(\Bk) = f_i^E(\Bk) + f_i^{\rm geo}(\Bk)$, with
\eqa{
& f_i^{E}(\Bk) = \ii \gamma \sum_n \del_{k_i} E_n(\Bk) P_n(\Bk)\ , \\
& f_i^{\rm geo}(\Bk) = \ii \gamma \sum_n E_n(\Bk) \del_{k_i} P_n(\Bk)\ . \\
\label{Pap_EQ:Splitting f_i}
}

\subsection{Different Contributions to EPC Constant \label{Appendix: Simplifying EPC}}

From Ref.\,\cite{mcmillan_transition_1968}, we know the EPC constant, derived via the Eliashberg function \cite{eliashberg_interactions_1960}, takes the form
\eqa{
\lambda = \frac{2 D(\mu)}{\hbar N} \frac{\left\langle \Gamma \right\rangle}{\langle \omega^2 \rangle}\ ,
\label{Pap_EQ:General EPC Expression}
}
where
\eqa{
\left\langle \Gamma \right\rangle = \frac{\sum_{\Bk,\Bk'}^{1 \text{BZ}} \sum_{nm} \Gamma_{nm}(\Bk,\Bk') S_{nm}(\Bk, \Bk')}{\sum_{\Bk,\Bk'}^{1 \text{BZ}} \sum_{nm} S_{nm}(\Bk, \Bk')}\ .
\label{Pap_EQ:General Expec Gamma Expression}
}
Here,
\eqa{
\Gamma_{nm}(\Bk,\Bk') = \frac{\hbar}{2} \sum_{\Btau,i} \frac{1}{m_{\Btau}} \Tr \lp P_n(\Bk) \left[ \chi_{\Btau} f_i(\Bk') - f_i(\Bk) \chi_{\Btau} \right] P_{m}(\Bk') \left[ \chi_{\Btau} f_i(\Bk) - f_i(\Bk') \chi_{\Btau} \right] \rp\ ,
\label{Pap_EQ:General Gamma Expression}
}
and
\eqa{
S_{nm}(\Bk, \Bk') = \delta(\mu - E_n(\Bk)) \delta(\mu - E_m(\Bk'))\ .
}
We note that even if we restore the spin degeneracy, the expression \cref{Pap_EQ:General EPC Expression} will not change in the absence of spin-orbit coupling.
In \cref{Pap_EQ:General EPC Expression}, $\langle \omega^2 \rangle$ is the mean-squared phonon frequency \cite{mcmillan_transition_1968}, $N$ the number of unit cells, $D(\mu)$ is the single particle density of states, and $\mu$ the chemical potential. In \cref{Pap_EQ:General Gamma Expression}, $m_{\Btau}$ is the mass of the atom at the sublattice $\Btau$. 1BZ denotes the first Brillouin zone. Since $f_i(\Bk) = f_i^E(\Bk) + f_i^{\rm geo}(\Bk)$, we may split \cref{Pap_EQ:General Gamma Expression}, and therefore \cref{Pap_EQ:General Expec Gamma Expression}, as:
\eqa{
\Gamma_{nm}^{X_1,X_2}(\Bk,\Bk') = \frac{\hbar}{2 m} \sum_{\Btau,i} \Tr \lp P_n(\Bk) \left[ \chi_{\Btau} f_i^{X_1}(\Bk') - f_i^{X_1}(\Bk) \chi_{\Btau} \right] P_{m}(\Bk') \left[ \chi_{\Btau} f_i^{X_2}(\Bk) - f_i^{X_2}(\Bk') \chi_{\Btau} \right] \rp 
}
where $X_1, X_2 = E,\mathrm{geo}$. As such, we find
\eqa{
& \la_E = \frac{2 D(\mu)}{\hbar N} \frac{\left\langle \Gamma \right\rangle^{E,E}}{\langle \omega^2 \rangle}\ , \\
& \la_{\rm geo} = \frac{2 D(\mu)}{\hbar N} \frac{\left\langle \Gamma \right\rangle^{\rm geo,geo}}{\langle \omega^2 \rangle}\ , \\
& \la_{E,\rm geo} = \frac{2 D(\mu)}{\hbar N} \frac{1}{\langle \omega^2 \rangle} \lp \left\langle \Gamma \right\rangle^{E, \rm geo} + \left\langle \Gamma \right\rangle^{\mathrm{geo}, E} \rp\ , \\
}
using
\eqa{
\left\langle \Gamma \right\rangle^{X_1, X_2} = \frac{\sum_{\Bk,\Bk'}^{1 \text{BZ}} \sum_{nm} \Gamma_{nm}^{X_1, X_2}(\Bk,\Bk') S_{nm}(\Bk, \Bk')}{\sum_{\Bk,\Bk'}^{1 \text{BZ}} \sum_{nm} S_{nm}(\Bk, \Bk')}\ , \quad X_1, X_2 = E, \rm geo\ .
}
We will handle each component separately. The following derivation holds for both the checkerboard model as well as the hyper-diamond model, in which case $t_{\Btau \Btau'}(\Br)$ is reduced to $t(\Br)$.

\subsection{Simplifying $\la$}

We now simplify the individual contributions to $\la$ generally. For $\la_{\rm geo}$, we have that
\eqa{
\left\langle \Gamma \right\rangle^{\rm geo,geo} = \frac{\sum_{\Bk,\Bk'}^{1 \text{BZ}} \sum_{nm} \Gamma_{nm}^{\rm geo,geo}(\Bk,\Bk') S_{nm}(\Bk, \Bk')}{\sum_{\Bk,\Bk'}^{1 \text{BZ}} \sum_{nm} S_{nm}(\Bk, \Bk')}\ .
\label{Pap_EQ:Gamma Geo-Geo Expec}
}
By inserting the expression for $\Gamma_{nm}^{\rm geo,geo}(\Bk,\Bk')$ into \cref{Pap_EQ:Gamma Geo-Geo Expec}, we may split $\left\langle \Gamma \right\rangle^{\rm geo,geo}$ into two terms:
\eqa{
& \left\langle \Gamma \right\rangle^{\rm geo,geo}_1 = -\frac{\hbar}{D^2(\mu)} \sum_{\Btau,i} \frac{1}{m_{\Btau}} \sum_{\Bk_1, \Bk_2} \sum_{n,m} \delta(\mu-E_n(\Bk_1)) \delta(\mu-E_m(\Bk_2)) \Tr \left\{ f_i^{\rm geo}(\Bk_1) P_n(\Bk_1) f_i^{\rm geo}(\Bk_1) \chi_{\Btau} P_m(\Bk_2) \chi_{\Btau} \right\}\ , \\
& \left\langle \Gamma \right\rangle^{\rm geo,geo}_2 = \frac{1}{D^2(\mu)} \frac{\hbar}{2} \sum_{\Btau,i} \frac{1}{m_{\Btau}} \Tr \left\{ \left( \sum_{\Bk_1} \sum_{n} \delta(\mu-E_n(\Bk_1)) \chi_{\Btau} f_i^{\rm geo}(\Bk_1) P_n(\Bk_1) \right)^2 + c.c. \right\}\ . \\
}
We define the term 
\eqa{
A_{i,\Btau} \equiv \sum_{\Bk} \sum_n \delta(\mu-E_n(\Bk)) \chi_{\Btau} f_i^{\rm geo}(\Bk) P_n(\Bk)\ ,
}
such that
\eqa{
\left\langle \Gamma \right\rangle^{\rm geo,geo}_2 = \frac{1}{D^2(\mu)} \frac{\hbar}{2} \sum_{\Btau,i} \frac{1}{m_{\Btau}} \Tr \left\{A_{i,\Btau}^2 + c.c. \right\}\ .
\label{Pap_EQ:Gamma Geo-Geo2 in terms of Ai}
}
We explore how a symmetry generator $\hat{g}$ acts on this term in \cref{Pap_SS: EPC Under Model Symmetries}. For now, we focus on $\left\langle \Gamma \right\rangle^{\rm geo,geo}_1$. By inserting the definition $f_i^{\rm geo}(\Bk)$ from \cref{Pap_EQ:Splitting f_i} into the definition of $\left\langle \Gamma \right\rangle^{\rm geo,geo}_1$, we find
\eqa{
\left\langle \Gamma \right\rangle^{\rm geo,geo}_1  = \frac{\gamma^2 \hbar}{D(\mu)} \sum_i \sum_{\Bk} \sum_{n,n_1,n_2} \delta(\mu-E_n(\Bk)) E_{n_1}(\Bk) E_{n_2}(\Bk) \Tr \left\{ \del_{k_i} P_{n_1}(\Bk) P_n(\Bk) \del_{k_i} P_{n_2}(\Bk) M \right\}\ ,
}
where
\eqa{
\label{eq:M_and_a}
& M = \frac{1}{D(\mu)} \sum_{\Btau} \frac{1}{m_{\Btau}} \sum_m \sum_{\Bk_2} \delta(\mu-E_m(\Bk_2)) \chi_{\Btau} P_m(\Bk_2) \chi_{\Btau} = \sum_{\Btau} a_{\Btau} \chi_{\Btau}\ , \\
& a_{\Btau} = \frac{1}{D(\mu)} \frac{1}{m_{\Btau}} \sum_m \sum_{\Bk_2} \delta(\mu-E_m(\Bk_2)) [P_m(\Bk_2)]_{\Btau \Btau}\ . \\
}
Since both models we consider have only two bands (see \cref{Appendix: Geo Enhancement in CB,Appendix: hyper-diamond Lattice}), we may sum over $n_1$ and $n_2$ to find
\eqa{
\left\langle \Gamma \right\rangle^{\rm geo,geo}_1  = \frac{\hbar \gamma^2}{D(\mu)} \sum_i \sum_{\Bk} \sum_n \delta(\mu-E_n(\Bk)) \Delta E^2(\Bk) \Tr \left\{ \del_{k_i} P_n(\Bk) P_n(\Bk) \del_{k_i} P_n(\Bk) M \right\}\ .
\label{Pap_EQ:Gamma geo-geo1 Unsimplified}
}
with
\eqa{
\Delta E^2(\Bk) = |E_2(\Bk) - E_1(\Bk)|^2\ ,
}
where we choose $E_1(\Bk) < E_2(\Bk)$, without loss of generality (WLOG). We can further simplify \cref{Pap_EQ:Gamma geo-geo1 Unsimplified} using parity $\P$ and time-reversal $\T$ symmetries, which we describe in \cref{Pap_SS: EPC Under Model Symmetries}. For now, we move on to $\la_E$ and $\la_{E,\rm geo}$.

In a similar method to the case of $\lambda_{\rm geo}$, we may insert the form of $f_i^E(\Bk)$ into the definition of $\Gamma_{nm}^{E,E}(\Bk,\Bk')$, therefore splitting $\left\langle \Gamma \right\rangle^{E,E}$ into two terms:
\eqa{
& \left\langle \Gamma \right\rangle^{E,E}_1 = -\frac{\hbar}{D^2(\mu)} \sum_{\Btau,i} \frac{1}{m_{\Btau}} \sum_{\Bk_1, \Bk_2} \sum_{n,m} \delta(\mu-E_n(\Bk_1)) \delta(\mu-E_m(\Bk_2)) \Tr \left\{ f_i^E(\Bk_1) P_n(\Bk_1) f_i^E(\Bk_1) \chi_{\Btau} P_m(\Bk_2) \chi_{\Btau} \right\}\ , \\
& \left\langle \Gamma \right\rangle^{E,E}_2 = \frac{1}{D^2(\mu)} \frac{\hbar}{2} \sum_{\Btau,i} \frac{1}{m_{\Btau}} \Tr \left\{ \left( \sum_{\Bk_1} \sum_{n} \delta(\mu-E_n(\Bk_1)) \chi_{\Btau} f_i^E(\Bk_1) P_n(\Bk_1) \right)^2 + c.c. \right\}\ . \\
\label{Pap_EQ:Gamma E-E Split}
}
We define the term 
\eqa{
B_{i,\Btau} \equiv \sum_{\Bk} \sum_n \delta(\mu-E_n(\Bk)) \chi_{\Btau} f_i^E(\Bk) P_n(\Bk)
}
such that
\eqa{
\left\langle \Gamma \right\rangle^{E,E}_2 = \frac{1}{D^2(\mu)} \frac{\hbar}{2} \sum_{\Btau,i} \frac{1}{m_{\Btau}} \Tr \left\{ B_{i,\Btau}^2 + c.c. \right\}\ .
\label{Pap_EQ:Gamma E-E2 in terms of Bi}
}
In \cref{Pap_SS: EPC Under Model Symmetries}, we explore the symmetry properties of $B_{i,\Btau}$, as well as further simplification of $\left\langle \Gamma \right\rangle^{E,E}_1$ due to symmetry.

Finally, we analyze $\la_{E,\rm geo}$. We have that
\eqa{
\la_{E, \rm geo} = \frac{2 D(\mu)}{\hbar N} \frac{1}{\langle \omega^2 \rangle} \lp \left\langle\Gamma\right\rangle^{E,\rm geo} + \left\langle\Gamma\right\rangle^{\mathrm{geo},E} \rp\ ,
}
with
\eqa{
& \left\langle\Gamma\right\rangle^{E, \rm geo} = \frac{ \sum_{\Bk_1, \Bk_2} \sum_{nm} \Gamma_{n m}^{E, \rm geo}(\Bk_1,\Bk_2) S_{nm}(\Bk_1, \Bk_2)}{\sum_{\Bk_1, \Bk_2} \sum_{nm} S_{nm}(\Bk_1, \Bk_2)}\ , \\
}
where
\eqa{
\Gamma_{n m}^{E,\rm geo}(\Bk,\Bk') = \frac{\hbar}{2} \sum_{\Btau,i} \frac{1}{m_{\Btau}} \Tr \left\{ P_n(\Bk) \left[\chi_{\Btau} f_i^{E}(\Bk') - f_i^{E}(\Bk) \chi_{\Btau} \right] P_{m}(\Bk') \left[\chi_{\Btau} f_i^{\rm geo}(\Bk) - f_i^{\rm geo}(\Bk') \chi_{\Btau} \right] \right\}\ .
}
$\left\langle\Gamma\right\rangle^{\mathrm{geo}, E}$ is the complex conjugate of $\left\langle\Gamma\right\rangle^{E,\rm geo}$. Further derivations are performed for $\left\langle\Gamma\right\rangle^{E,\rm geo}$, however they hold for $\left\langle\Gamma\right\rangle^{\mathrm{geo}, E}$ as well. We insert \cref{Pap_EQ:Splitting f_i}, and, using
\eqa{
f_i^{\rm geo}(\Bk) P_n(\Bk) = \ii \gamma \sum_m E_m(\Bk) \del_{k_i} P_m(\Bk) P_n(\Bk) = \ii \gamma (-1)^n \Delta E(\Bk) \del_{k_i} P_n(\Bk) P_n(\Bk)
}
as $\del_{k_i} P_m(\Bk) = - \del_{k_i} P_n(\Bk)$ for $m \neq n$ in a two-band model, we find
\eqa{
\Gamma_{n n'}^{E,\rm geo}(\Bk,\Bk') & = -\gamma^2 \frac{\hbar}{2} \sum_{\Btau,i} \frac{1}{m_{\Btau}} \left[ \del_{k_i'} E_{n'}(\Bk') - \del_{k_i}E_n(\Bk) \right] \Tr \left\{ P_n(\Bk) \chi_{\Btau} P_{n'}(\Bk') \chi_{\Btau} (-1)^n \Delta E(\Bk) \del_{k_i} P_n(\Bk) \right\} \\
&\quad + \gamma^2 \frac{\hbar}{2} \sum_{\Btau,i} \frac{1}{m_{\Btau}} \left[ \del_{k_i'} E_{n'}(\Bk') - \del_{k_i}E_n(\Bk) \right] \Tr \left\{ P_n(\Bk) \chi_{\Btau} P_{n'}(\Bk') (-1)^{n'} \Delta E(\Bk') \del_{k_i'} P_{n'}(\Bk') \chi_{\Btau} \right\} \\
& = -\gamma^2 \frac{\hbar}{2} \sum_{\Btau,i} \frac{1}{m_{\Btau}} \left[ \del_{k_i'} E_{n'}(\Bk') - \del_{k_i}E_n(\Bk) \right] (-1)^n \Delta E(\Bk) \mathcal{A}_{\Btau,n,i}(\Bk) \Tr \left[ P_{n'}(\Bk') \chi_{\Btau} \right] \\
&\quad + \gamma^2 \frac{\hbar}{2} \sum_{\Btau,i} \frac{1}{m_{\Btau}} \left[ \del_{k_i'} E_{n'}(\Bk') - \del_{k_i}E_n(\Bk) \right] (-1)^{n'} \Delta E(\Bk') \mathcal{A}^*_{\Btau,n',i}(\Bk') \Tr \left[ P_{n}(\Bk) \chi_{\Btau} \right]\ , \\
& = -\gamma^2 \frac{\hbar}{2} \sum_{\Btau,i} \frac{1}{m_{\Btau}}  \del_{k_i'} E_{n'}(\Bk')(-1)^n \Delta E(\Bk) \mathcal{A}_{\Btau,n,i}(\Bk) \Tr \left[ P_{n'}(\Bk') \chi_{\Btau} \right] \\
& \quad  +\gamma^2 \frac{\hbar}{2} \sum_{\Btau,i} \frac{1}{m_{\Btau}}   \del_{k_i}E_n(\Bk) (-1)^n \Delta E(\Bk) \mathcal{A}_{\Btau,n,i}(\Bk) \Tr \left[ P_{n'}(\Bk') \chi_{\Btau} \right] \\
&\quad + \gamma^2 \frac{\hbar}{2} \sum_{\Btau,i} \frac{1}{m_{\Btau}}  \del_{k_i'} E_{n'}(\Bk') (-1)^{n'} \Delta E(\Bk') \mathcal{A}^*_{\Btau,n',i}(\Bk') \Tr \left[ P_{n}(\Bk) \chi_{\Btau} \right]\\
&\quad  - \gamma^2 \frac{\hbar}{2} \sum_{\Btau,i} \frac{1}{m_{\Btau}}  \del_{k_i}E_n(\Bk)  (-1)^{n'} \Delta E(\Bk') \mathcal{A}^*_{\Btau,n',i}(\Bk') \Tr \left[ P_{n}(\Bk) \chi_{\Btau} \right]\ , 
\label{Pap_EQ:Gamme E-Geo in terms of Qi}
}
where
\eqa{
\mathcal{A}_{\Btau,n,i}(\Bk) = \Tr \left[ \chi_{\Btau} \del_{k_i} P_n(\Bk) P_n(\Bk) \right]
\label{eq:A_field_def}
}
is an orbital-selective complex vector field. 
Then, 
\eqa{
\left\langle\Gamma\right\rangle^{E,\rm geo} & = \gamma^2 \hbar\frac{1}{D^2(\mu)} \sum_{n,n'} \sum_{\Btau,i}\frac{1}{m_{\Btau}} \left[  - (-1)^n     \sum_{\Bk'} \delta(\mu-E_{n'}(\Bk'))\del_{k_i'} E_{n'}(\Bk') \Tr \left[ P_{n'}(\Bk') \chi_{\Btau} \right]  \sum_{\Bk}\delta(\mu-E_n(\Bk)) \Delta E(\Bk) \Re(\mathcal{A}_{\Btau,n,i}(\Bk)) \right. \\
& \quad  \left. +    \sum_{\Bk} \delta(\mu-E_{n}(\Bk)) \del_{k_i}E_n(\Bk) (-1)^n \Delta E(\Bk) \Re(\mathcal{A}_{\Btau,n,i}(\Bk)) \sum_{\Bk'} \delta(\mu-E_{n'}(\Bk')) \Tr \left[ P_{n'}(\Bk') \chi_{\Btau} \right]\right]\\
& = \ii \gamma \hbar\frac{1}{D^2(\mu)} \sum_{n}  (-1)^n \sum_{\Btau,i}     \frac{1}{m_{\Btau}} B_{i,\Btau} \sum_{\Bk}\delta(\mu-E_n(\Bk)) \Delta E(\Bk) \Re(\mathcal{A}_{\Btau,n,i}(\Bk))\\
& \quad + \gamma^2 \hbar\frac{1}{D(\mu)} \sum_{n} \sum_{\Btau,i} \sum_{\Bk} \delta(\mu-E_{n}(\Bk)) \del_{k_i}E_n(\Bk) (-1)^n \Delta E(\Bk) \Re(\mathcal{A}_{\Btau,n,i}(\Bk)) a_{\Btau}  \\
& = \left\langle\Gamma\right\rangle^{E, \rm geo}_1 + \left\langle\Gamma\right\rangle^{E,\rm geo}_2 \ ,\\
}
where
\eqa{
\label{eq:E-geo}
& \left\langle\Gamma\right\rangle^{E,\rm geo}_1 = \gamma^2 \hbar\frac{1}{D(\mu)} \sum_{n} \sum_{\Btau,i}\sum_{\Bk} \delta(\mu-E_{n}(\Bk)) \del_{k_i}E_n(\Bk) (-1)^n \Delta E(\Bk) \Re(\mathcal{A}_{\Btau,n,i}(\Bk)) a_{\Btau}\ , \\
& \left\langle\Gamma\right\rangle^{E,\rm geo}_2 = \ii \gamma \hbar\frac{1}{D^2(\mu)} \sum_{n}  (-1)^n \sum_{\Btau,i}     \frac{1}{m_{\Btau}} B_{i,\Btau} \sum_{\Bk}\delta(\mu-E_n(\Bk)) \Delta E(\Bk) \Re(\mathcal{A}_{\Btau,n,i}(\Bk))\ . \\
}
Next we further simplify the different contributions to $\la$ using symmetry constraints of the checkerboard and hyper-diamond models.

\subsection{Further Simplification of $\la$ Derived from Model Symmetries \label{Pap_SS: EPC Under Model Symmetries}}

We further simplify the different contributions to $\la$ using arguments derived from symmetry. In both the checkerboard and hyper-diamond models, we assume two spinless orbitals per unit cell. We here describe a general formalism for symmetries in solids, and will in particular discuss the $\P \T$ symmetry that is present in both models, where $\P$ is inversion and $\T$ is time-reversal. 

\subsubsection{Representations of Symmetries \label{Pap_SS:Symmetry Reps}}

We first describe how symmetries act on our basis. This applies to both the checkerboard model and hyper-diamond model discussed in \cref{Appendix: hyper-diamond Lattice,Appendix: Geo Enhancement in CB}. Suppose $\hat{g}$ is a symmetry representation of the space group, and $\mathcal{G}$ is the group of the model. Let $\rmg$ be a symmetry operation in $\G$, such that $\rmg = \{ R | \Bd \} \in \G$, where $R$ is the rotational part and $\Bd$ is the translational part of the operation. Then
\eqa{
\hat{g} c_{\BR+\Btau}^\dagger \hat{g}^{-1} = c_{\BR_{\Btau,g}+\Btau_g}^\dagger\ ,
\label{Pap_EQ:g on real-space creation}
}
where $\BR_{\Btau,g}$ and $\Btau_g$ satisfy
\eqa{
& \BR_{\Btau,g} = R \BR + \Delta \BR_{\Btau,g}\ , \\
& R \Btau + \Bd = \Btau_g + \Delta \BR_{\Btau,g}\ , \\
\label{Pap_EQ:Rg and taug Conditions}
}
with $\Delta \BR_{\Btau,g}$ a lattice vector.
Then $\rmg$ acting on $\BR+\Btau$ has the form
\eqa{
R(\BR + \Btau) + \Bd & = R\BR + R\Btau + \Bd \\
& = R\BR + \Btau_g + \Delta \BR_{\Btau,g} \\
& = \BR_{\Btau,g} + \Btau_g\ . \\
}
For time-reversal symmetry $\T$, we have
\eqa{
\T c_{\BR+\Btau}^\dagger \T^{-1} = c_{\BR+\Btau}^\dagger\ .
\label{Pap_EQ:TR on real-space creation}
}



%
Recall that $R\Btau+\Bd = \Btau_g + \Delta \BR_{\Btau,g}$ with $\Delta \BR_{\Btau,g}$ a lattice vector, using \cref{Pap_EQ:Rg and taug Conditions}. We also have that $\hat{g} c_{\BR+\Btau}^\dagger \hat{g}^{-1} = c_{R\BR+R\Btau +\Bd}^\dagger$. Using \cref{Pap_EQ:FT Rule},
\eqa{
\hat{g} c_{\Bk,\Btau}^\dagger \hat{g}^{-1} & = \frac{1}{\sqrt{N}} \sum_{\BR} e^{\ii \Bk \cdot (\BR+\Btau)} \hat{g} c_{\BR+\Btau}^\dagger \hat{g}^{-1} \\
& = \frac{1}{\sqrt{N}} \sum_{\BR} e^{\ii \Bk \cdot (\BR+\Btau)} c^\dagger_{R\BR+R\Btau+\Bd} \\
& = \frac{1}{\sqrt{N}} \sum_{\BR} e^{\ii \Bk \cdot (\BR+\Btau)} c_{R \BR+\Btau_g+\Delta\BR_{\Btau,g}}^\dagger\ . \\
}
We define $\BR' \equiv R\BR+\Delta\BR_{\Btau,g}$. Then
\eqa{
\hat{g} c_{\Bk,\Btau}^\dagger \hat{g}^{-1} & = \frac{1}{\sqrt{N}} \sum_{\BR} e^{\ii \Bk \cdot (\BR+\Btau)} c_{\BR'+\Btau_g}^\dagger \\
& = \frac{1}{\sqrt{N}} \sum_{\BR} e^{\ii \Bk \cdot (\BR+\Btau)} \left[
\frac{1}{\sqrt{N}} \sum_{\Bk'} e^{-\ii \Bk' \cdot (\BR'+\Btau_g)} c_{\Bk', \Btau_g}^\dagger \right] \\
& = \frac{1}{N} \sum_{\BR} \sum_{\Bk'} e^{\ii \Bk \cdot (\BR + \Btau)} e^{-\ii \Bk' \cdot (R\BR + \Btau_g + \Delta \BR_{\Btau,g})} c_{\Bk',\Btau_g}^\dagger\ . \\
}
Observe $\Bk' \cdot R\BR = R^T \Bk' \cdot \BR = R^{-1} \Bk' \cdot \BR$, since $R$ is an orthonormal rotation matrix. Then 
\eqa{
\hat{g} c_{\Bk,\Btau}^\dagger \hat{g}^{-1} & = \frac{1}{N} \sum_{\Bk'} e^{\ii \Bk \cdot \Btau} e^{-\ii \Bk' \cdot (\Btau_g + \Delta \BR_{\Btau,g})} \left[ 
\sum_\BR e^{\ii \BR \cdot (\Bk - R^{-1}\Bk')} \right] c_{\Bk',\Btau_g}^\dagger \\
& = \frac{1}{N} \sum_{\Bk'} e^{\ii \Bk \cdot \Btau} e^{-\ii \Bk' \cdot (\Btau_g + \Delta \BR_{\Btau,g})} \left[ N \delta_{\Bk,R^{-1}\Bk'} \right] c_{\Bk',\Btau_g}^\dagger \\
& = e^{\ii \Bk \cdot \Btau} e^{-\ii R\Bk \cdot (\Btau_g+\Delta\BR)} c_{R\Bk,\Btau_g}^\dagger \\
& = e^{-\ii R\Bk \cdot \Bd} c_{R\Bk,\Btau_g}^\dagger\ , \\
}
since $R\Btau +\Bd= \Btau_g + \Delta\BR$.

Define $U_{g}$ as $\left[ U_{g} \right]_{\Btau'\Btau} = \delta_{\Btau' \Btau_g}$, we arrive at
\eq{
 \hat{g} c_{\Bk}^\dagger \hat{g}^{-1} = e^{-\ii R\Bk \cdot \Bd} c_{R\Bk}^\dagger U_g\ ,
}
with $c_{\Bk}^\dagger$ having two components.

\subsubsection{Applying Symmetries to $\left\langle \Gamma \right\rangle^{\rm geo,geo}_2$, $\left\langle \Gamma \right\rangle^{E,E}_2$ and $\left\langle \Gamma \right\rangle^{E,\rm geo}_2$ }

We may now simplify the different contributions to $\la$ using symmetry constraints.
From now on, we set
\eq{
\Bd = 0
}
which means we only consider the point group symmetries. We also assert that $U_g$ acts on the sublattice as identity, \ie, $U_g$ is diagonal in the sublattice.

We begin by returning to $\left\langle \Gamma \right\rangle^{\rm geo,geo}_2$ in \cref{Pap_EQ:Gamma Geo-Geo2 in terms of Ai}. We ask how $A_{i,\Btau}$ transforms under a symmetry operation $U_g$. Suppose $\hat{g}$ is a symmetry representation of the space group. Observe $\Tr \left\{ U_g A_{i,\Btau}^2 U_g^{-1} \right\} = \Tr \left\{ A_{i,\Btau}^2 \right\}$ by the cyclic property of traces, but $\Tr \left\{ U_g A_{i,\Btau}^2 U_g^{-1} \right\} = \Tr \left\{ U_g A_{i,\Btau} A_{i,\Btau} U_g^{-1} \right\} = \Tr \left\{ U_g A_{i,\Btau} U_g^{-1} U_g A_{i,\Btau} U_g^{-1} \right\} = \Tr \left\{ (U_g A_{i,\Btau} U_g^{-1})^2 \right\}$. Thus $\Tr \left\{ A_{i,\Btau}^2 \right\} = \Tr \left\{ (U_g A_{i,\Btau} U_g^{-1})^2 \right\}$. Observe
\eqa{
U_g A_{i,\Btau} U_g^{-1} & = U_g \left[ \sum_{\Bk} \sum_n \delta(\mu-E_n(\Bk)) \chi_{\Btau} f_i^{\rm geo}(\Bk) P_n(\Bk) \right] U_g^{-1} \\
& = \sum_{\Bk} \sum_n \delta(\mu-E_n(\Bk)) \chi_{\Btau} U_g f_i^{\rm geo}(\Bk) U_g^{-1} U_g P_n(\Bk) U_g^{-1}\ . \\
}
Owing to $U_g h(\Bk) U_g^{-1} = h(R \Bk)$, as well as $\hat{g} \ket{\psi_n(\Bk)} = \ket{\psi_n(R \Bk)} e^{\ii \th(\Bk)}$ with $\ket{\psi_n(\Bk)} = \sum_{\Btau} c_{\Bk,\Btau}^\dagger [U_n(\Bk)]_{\Btau} \ket{0}$ the Bloch wavefunction,
we have $U_g P_n(\Bk) U_g^{-1} = P_n(R\Bk)$ and $E_n(\Bk) = E_n(R\Bk)$. If we define $\Bk' \equiv R\Bk$, we can show
\eqa{
\del_{k_i} \equiv \frac{\del}{\del k_i} = \sum_j \frac{\del k_j'}{\del k_i} \frac{\del}{\del k_j'} = \sum_j \frac{\del k_j'}{\del k_i} \del_{k_j'}\ .
}
Combined with
\eqa{
\frac{\del k_j'}{\del k_i} = \frac{\del}{\del k_i} \left( \sum_l R_{jl} k_l \right) = \sum_l R_{jl} \frac{\del k_l}{\del k_i} = \sum_l R_{jl} \delta_{li} = R_{ji}\ ,
}
we obtain $U_g f_i^{\rm geo}(\Bk) U_g^{-1} = \sum_j R_{ji} f_j^{\rm geo}(R\Bk)$. Thus,
\eqa{
U_g A_{i,\Btau} U_g^{-1} & = \sum_j R_{ji} \sum_{\Bk'} \sum_n \delta(\mu-E_n(\Bk')) \chi_{\Btau}f_j^{\rm geo}(\Bk') P_n(\Bk') \\
& = \sum_j R_{ji} A_{j,\Btau}\ . \\
}
Here we used the fact that $R$ maps the 1BZ onto itself, modulo reciprocal lattice vectors on measure-zero boundary.\
%
Since we assume $U_g$ acts on the sublattice as identity, we have $U_g = 1$, and we find that
\eqa{
A_{i,\Btau} = \sum_j R_{ji} A_{j,\Btau}\ .
}
Clearly, $A_{i,\Btau}$ with any fixed $\Btau$ furnish a trivial representation of $R$ for all point-group $\hat{g}$ with symmetry operations that act on $\Btau$ as identity.

We now discuss $\left\langle \Gamma \right\rangle^{E,E}_2$ in \cref{Pap_EQ:Gamma E-E2 in terms of Bi} and $\left\langle \Gamma \right\rangle^{E, \rm geo}_2$ in \cref{eq:E-geo}.
Similarly, for any point-group $\hat{g}$ with symmetry operations that act on the sublattice as identity, we have
\eqa{
U_g f_i^E(\Bk) U_g^{-1} & = U_g \left[ \ii \gamma \sum_n \del_{k_i} E_n(\Bk) P_n(\Bk) \right] U_g^{-1} \\
& = \ii \gamma \sum_n \del_{k_i} U_g E_n( \Bk) U_g^{-1} U_g P_n(\Bk) U_g^{-1} \\
& = \ii \gamma \sum_n \sum_j R_{ji} \del_{k_j} E_n(R\Bk) P_n(R\Bk) \\
& = \sum_j R_{ji} f_j^E(R\Bk)\ ,
}
which means
\eqa{
U_g B_{i,\Btau} U_g^{-1} = \sum_{\Bk} \sum_n \delta(\mu-E_n(\Bk)) \chi_{\Btau} f_i^E(\Bk) P_n(\Bk) = \sum_j R_{ji} B_{j,\Btau}\ .
}
Clearly, $B_{i,\Btau}$ with any fixed $\Btau$ furnish a trivial representation of $R$ for all point-group $\hat{g}$ with symmetry operations that act on $\Btau$ as identity.

Clearly, all $\left\langle \Gamma \right\rangle^{\rm geo,geo}_2$, $\left\langle \Gamma \right\rangle^{E,E}_2$ and $\left\langle \Gamma \right\rangle^{E,\rm geo}_2$ rely on quantities that are invariant under $R$ of all $\hat{g}$ with symmetry operations which act as identity on the sublattice.
If $R$ of all $\hat{g}$ with symmetry operations that act as identity on the sublattice do not allow such a representation, $\left\langle \Gamma \right\rangle^{\rm geo,geo}_2$, $\left\langle \Gamma \right\rangle^{E,E}_2$ and $\left\langle \Gamma \right\rangle^{E,\rm geo}_2$ must be zero.
This is actually the case for  both the checkerboard and hyper-diamond models, as discussed in \cref{Pap_SS:HypD Symmetries,Pap_SS:CB Symmetries} for details.

\subsubsection{Applying Symmetries to $\left\langle \Gamma \right\rangle^{\rm geo,geo}_1$, $\left\langle \Gamma \right\rangle^{E,E}_1$ and $\left\langle \Gamma \right\rangle^{E,\rm geo}_1$ }

We now discuss $\left\langle \Gamma \right\rangle^{\rm geo,geo}_1$, $\left\langle \Gamma \right\rangle^{E,E}_1$ and $\left\langle \Gamma \right\rangle^{E,\rm geo}_1$ based on the $\PT$ symmetry. Recall both models have two spinless orbitals per unit cell. If we choose the origin to be at the midpoint between the two orbitals such that $\Btau_1 = -\Btau_2$, we find inversion symmetry $\mathcal{P}$ at the origin to be a symmetry operation. Inversion maps $\P\BR_{\Btau} = -\BR$ and $\Btau_\P = - \Btau = \P \Btau$ which means $\Delta \BR_{\P,\Btau}=0$. Thus, from \cref{Pap_EQ:g on real-space creation},
\eqa{
\mathcal{P} c_{\BR + \Btau}^\dagger \mathcal{P}^{-1} = c_{\P\BR+\P\Btau}^\dagger = c_{-\BR -\Btau}^\dagger\ .
\label{Pap_EQ:Parity_on_Creation}
}
We apply \cref{Pap_EQ:FT Rule} and find $\P c_{\Bk,\Btau}^\dagger \P^{-1} = \frac{1}{\sqrt{N}} \sum_{\BR} e^{\ii \Bk \cdot (\BR+\Btau)} c_{-\BR-\Btau}^\dagger$. We insert $c_{-\BR-\Btau}^\dagger = \frac{1}{\sqrt{N}} \sum_{\Bk'} e^{-\ii \Bk' \cdot (-\BR-\Btau)} c_{\Bk',-\Btau}^\dagger$, and find $\P c_{\Bk,\Btau}^\dagger \P^{-1} = c_{-\Bk,-\Btau}^\dagger$. We perform a similar derivation for time-reversal symmetry $\T$, and find $\mathcal{T} c_{\Bk,\Btau}^\dagger \mathcal{T}^{-1} = c_{-\Bk,\Btau}^\dagger$. Thus
\eqa{
\mathcal{PT} c_{\Bk,\Btau}^\dagger (\mathcal{PT})^{-1} = c_{\Bk,-\Btau}^\dagger\ .
}
Thus $\mathcal{PT} c_{\Bk}^\dagger (\mathcal{PT})^{-1} = c_{\Bk}^\dagger \tau_x$. Then, if $\ket{\psi_n(\Bk)} = \sum_{\Btau} c_{\Bk,\Btau}^\dagger [U_n(\Bk)]_{\Btau} \ket{0}$, we find 
\eqa{
(\mathcal{PT}) \ket{\psi_n(\Bk)} = \sum_{\Btau} c^\dagger_{\Bk,-\Btau} [U_n(\Bk)]^*_{\Btau} \ket{0}\ . 
}
This implies that $\tau_x P_n^*(\Bk) \tau_x = P_n(\Bk)$, which further implies $[P_n(\Bk)]_{\Btau \Btau} = \frac{1}{2}$. Thus, $a_{\Btau} = \frac{1}{2 m}$ in \cref{eq:M_and_a}, with $m_{\Btau} = m$ for all sublattices owing to the $\PT$ symmetry. Observe that if $F(\Bk + \forall\BG) = F(\Bk)$, then
\eqa{
\sum_{\Bk \in 1\text{BZ}} \delta(\mu - E_n(\Bk)) F(\Bk) = \frac{\mathcal{V}}{(2 \pi)^d} \int\limits_{\textsf{FS}_n} d\sSigma_{n,\Bk} \frac{1}{\big|\Bnabla_\Bk E_n(\Bk)\big|} F(\Bk)\ ,
\label{Pap_EQ:Sum to Int Relation}
}
in the thermodynamic limit \cite{yu_nontrivial_2024}, where $\mathcal{V}$ is the volume of the system, $\textsf{FS}_n$ is the Fermi surface given by $E_n(\Bk) = \mu$, and $d\sSigma_{n,\Bk}$ is a volume element on $\textsf{FS}_n$. Since the system is periodic, we have 
\eqa{
\lambda_{\rm geo,1} = \frac{\gamma^2}{m \langle \omega^2 \rangle} \sum_n \frac{\sOmega}{(2\pi)^d} \int\limits_{\textsf{FS}_n} d\sSigma_{n,\Bk} \frac{\Delta E^2(\Bk)}{|\Bnabla_\Bk E_n(\Bk)|} \Tr [g_n(\Bk)]\ , \\
\label{Pap_EQ:Lambda Geo FS Form}
}
where $\sOmega$ is the volume of the unit cell, and $g_n(\Bk)$ is the quantum metric, given by
\eqa{
[g_n(\Bk)]_{ij} & = \frac{1}{2} \Tr \left[ \frac{\del P_n(\Bk)}{\del k_i} \frac{\del P_n(\Bk)}{\del k_j} \right] \\
& = \frac{1}{2} \Tr \left[ \frac{\del P_n(\Bk)}{\del k_i} P_n(\Bk) \frac{\del P_n(\Bk)}{\del k_j} \right] + (i \leftrightarrow j)\ . \\
\label{Pap_EQ:FSM Definition}
}
The simplification of $\left\langle \Gamma \right\rangle^{\rm geo,geo}_2 = 0$ relies on other symmetries in the checkerboard and hyper-diamond models.

We now move on to simplifying $\left\langle \Gamma \right\rangle^{E,E}_1$ as given in \cref{Pap_EQ:Gamma E-E Split}. By inserting $f_i^E(\Bk)$ from \cref{Pap_EQ:Splitting f_i} due to $\PT$ symmetry, we find
\eqa{
\left\langle \Gamma \right\rangle^{E,E}_1 = \frac{\gamma^2 \hbar}{D(\mu)} \sum_{\Bk} \sum_{\Btau, i} \frac{1}{2 m} \sum_{n, n_1, n_2} \delta(\mu-E_n(\Bk)) \del_{k_i} E_{n_1}(\Bk) \del_{k_i} E_{n_2}(\Bk) \Tr \left\{ P_{n_1}(\Bk) P_n(\Bk) P_{n_2}(\Bk) \chi_{\Btau} \right\}\ .
}
We define the term
\eqa{
F_{n,\Btau}(\Bk) = P_{1}(\Bk) P_n(\Bk) P_{1}(\Bk) \chi_{\Btau} + P_{2}(\Bk) P_n(\Bk) P_{2}(\Bk) \chi_{\Btau} - 2 P_{1}(\Bk) P_n(\Bk) P_{2}(\Bk) \chi_{\Btau}\ ,
}
where $n=1,2$. We can show that
\eqa{
\Tr \left\{ F_{n,\Btau}(\Bk) \right\} = [P_n(\Bk)]_{\Btau \Btau}\ .
}
As such,
\eqa{
\sum_{\Btau} \Tr \left\{ F_{n,\Btau}(\Bk) \right\} = \Tr \left\{ P_n(\Bk) \right\} = 1\ .
}
Thus, the sum becomes
\eqa{
& \sum_{\Btau,i} \frac{1}{2 m_{\Btau}} \sum_{n, n_1, n_2} \delta(\mu-E_n(\Bk)) \del_{k_i} E_{n_1}(\Bk) \del_{k_i} E_{n_2}(\Bk) \Tr \left\{ P_{n_1}(\Bk) P_n(\Bk) P_{n_2}(\Bk) \chi_{\Btau} \right\} \\
& = \sum_{\Btau,i} \frac{1}{2 m_{\Btau}} \delta(\mu-E_1(\Bk)) |\del_{k_i}E_1(\Bk)|^2 \Tr \left\{ F_{1,\Btau}(\Bk) \right\} + \sum_{\Btau,i} \frac{1}{2 m_{\Btau}} \delta(\mu-E_2(\Bk)) |\del_{k_i}E_2(\Bk)|^2 \Tr \left\{ F_{2,\Btau}(\Bk) \right\} \\
& = \frac{1}{2 m} \sum_{i} \delta(\mu-E_1(\Bk)) |\del_{k_i}E_1(\Bk)|^2 + \frac{1}{2 m} \sum_{i} \delta(\mu-E_2(\Bk)) |\del_{k_i}E_2(\Bk)|^2 \\
& = \frac{1}{2 m} \sum_{n, i} \delta(\mu-E_n(\Bk)) |\del_{k_i}E_n(\Bk)|^2 \\
& = \frac{1}{2 m} \sum_n \delta(\mu-E_n(\Bk)) | \Bnabla_{\Bk} E_n(\Bk)|^2\ , \\
}
again assuming $m_{\Btau} = m$ for all sublattices. As such, we have
\eqa{
\left\langle \Gamma \right\rangle^{E,E}_1 = \frac{\gamma^2}{m D(\mu)} \sum_{\Bk} \frac{\hbar}{2} \sum_n \delta(\mu-E_n(\Bk)) \big| \Bnabla_{\Bk} E_n(\Bk) \big|^2\ .
}
Again employing \cref{Pap_EQ:Sum to Int Relation}, we find
\eqa{
\lambda_{E,1} = \frac{\gamma^2}{m \langle \omega^2 \rangle} \sum_n \frac{\sOmega}{(2\pi)^d} \int\limits_{\textsf{FS}_n} d\sSigma_{n,\Bk} \big| \Bnabla_{\Bk} E_n(\Bk) \big|\ .
\label{Pap_EQ:Lambda E FS Form}
}

At last, let's look at $\left\langle \Gamma \right\rangle^{E,\rm geo}_1$ in \cref{eq:E-geo}.
Owing to $a_{\Btau}=1/(2m)$, we have
\eq{
\left\langle\Gamma\right\rangle^{E, \rm geo}_1 = \gamma^2 \hbar\frac{1}{2 m D(\mu)} \sum_{n} \sum_{i}\sum_{\Bk} \delta(\mu-E_{n}(\Bk)) \del_{k_i}E_n(\Bk) (-1)^n \Delta E(\Bk) \sum_{\Btau}\Re(\mathcal{A}_{\Btau,n,i}(\Bk)) \ .
}
As 
\eq{
\sum_{\Btau}\mathcal{A}_{\Btau,n,i}(\Bk) =  \Tr \left[ \del_{k_i} P_n(\Bk) P_n(\Bk) \right] = 0\ ,
}
we know $\left\langle\Gamma\right\rangle^{E,\rm geo}_1 =0$.

\section{Geometric Contribution to EPC in the Checkerboard Model \label{Appendix: Geo Enhancement in CB}}

\subsection{Checkerboard Model Symmetry Properties \label{Pap_SS:CB Symmetries}}

We now describe the process of solving $W_{i,\Btau} = \sum_j R_{ji} W_{j,\Btau}$ in the case of the checkerboard model \cite{sun_time-reversal_2008, sun_topological_2009, li_magic-angle_2022}.
We again begin with \cref{Pap_EQ:Tight-Binding Hamiltonian}. 
As shown in \cref{Main_Fig:CB_plots}, there are two sublattices $A$ and $B$, with sublattice $A$ at $\Btau_1 = -\frac{1}{4} (\Ba_1 + \Ba_2)$ and sublattice $B$ at $\Btau_2 = \frac{1}{4} (\Ba_1 + \Ba_2)$, where $\Ba_i = a \hat{\bsl{e}}_i$ are the primitive lattice vectors. We again choose each atom to have one spinless orbital, and we set the lattice constant $a=1$ for convenience. We use a tight-binding model \cref{Pap_EQ:Tight-Binding Hamiltonian}, convert to reciprocal space using \cref{Pap_EQ:FT Rule}, and find the Hamiltonian has the form $H = \sum_\Bk c_\Bk^\dagger h(\Bk) c_\Bk$, where
\eqa{
h(\Bk) = d_x(\Bk) \sigma^x + d_z(\Bk) \sigma^z\ ,
}
with
\eqa{
& d_x(\Bk) = 4 t \cos \left( \frac{k_x}{2} \right) \cos \left( \frac{k_y}{2} \right) \\
& d_z(\Bk) = -2 t' \left[ \cos(k_x) - \cos(k_y) \right]\ , \\
}
where $t$ and $t'$ are hopping parameters corresponding to nearest and next-nearest-neighbor hoppings, respectively, and $c_\Bk^\dagger = \lp c_{\Bk,\Btau_1}^\dagger,c_{\Bk,\Btau_2}^\dagger \rp$ \cite{li_magic-angle_2022}. We show the resulting band structure for $t' = t/2$ in \cref{Main_Fig:CB_plots}. In our analysis, we chose the inversion center $\BR_0 = \frac{1}{4}(\hat{\bsl{e}}_1 + \hat{\bsl{e}}_2)$. The checkerboard model has point group symmetry $D_{4h}$. If a symmetry operation $\rmg$ acts on the sublattice as identity, \ie, $R \Btau_i + (I-R) \BR_0 = \Btau_i + \BR$, where $\BR$ is a lattice vector, then we generate a set of equations $W_{i,\Btau} = \sum_j R_{ji} W_{j,\Btau}$ using the point group representations from \refcite{aroyo_bilbao_2006}, which can be seen in \cref{Pap_EQ:Checkerboard Symmetry Derivation}.
\begin{subequations}
\eqa{
C_{2z} : & \quad W_{1,\Btau} = - W_{1,\Btau} \\
& \quad W_{2,\Btau} = - W_{2,\Btau} \\
}
\eqa{
C_{2,110} : \quad W_{3,\Btau} = - W_{3,\Btau}
}
\label{Pap_EQ:Checkerboard Symmetry Derivation}
\end{subequations}
Thus, $W_{i,\Btau} = 0$ for $i = 1, 2, 3$ in the checkerboard model.
In other words, $C_{2z}$ and $C_{2,110}$ already do not admit invariant representation.

\subsection{Geometric Enhancement \label{Pap_SS:Geo Enhancement}}

We now show the enhancement of $\la_{\rm geo}$ in the checkerboard model. 
We analyze $\la_E$ and $\la_{\rm geo}$ when $\mu$ is close to the $\Gamma$ point dispersion. In this region, we can show that the geometric contribution, given by \cref{Pap_EQ:Lambda Geo FS Form}, can be tuned by varying the hopping parameter $t'$, while the energetic contribution \cref{Pap_EQ:Lambda E FS Form} is constant, when considering lowest order contributions. We assume $\mu < 0$ and $t, t' > 0$.
In this region, the Fermi surface is set by $\textsf{FS}_\pm: \mu = E_\pm(\Gamma)$. For a two-band Hamiltonian of the form $h(\Bk) = \Bd(\Bk) \cdot \mbf{\sigma}$, where $\mbf{\sigma} = (\sigma^x, \sigma^y, \sigma^z)$ is the vector of Pauli matrices, we find $E_\pm(\Bk) = \pm | \Bd(\Bk) |$ and $P_\pm(\Bk) = \frac{1}{2} \sigma^0 \pm \frac{1}{2} \hat{\Bd}(\Bk) \cdot \mbf{\sigma}$, where $\Bd(\Bk) = \sum_i d_i(\Bk) \hat{\bsl{e}}_i$, $\hat{\Bd}(\Bk) \equiv \Bd(\Bk) / |\Bd(\Bk)|$, and $\sigma^0$ is the identity matrix. In general, we know $\Delta E(\Bk) = 2 |\Bd(\Bk)|$ and $\del_{k_i} P_\pm(\Bk) = \pm \frac{\mbf{\sigma}}{2} \cdot \del_{k_i} \hat{\Bd}(\Bk)$. Observe,
\eqa{
\del_{k_i} \hat{\Bd}(\Bk) & = \del_{k_i} \frac{\Bd(\Bk)}{\big| \Bd(\Bk) \big|} \\
& = \frac{\del_{k_i} \Bd(\Bk)}{\big| \Bd(\Bk) \big|} + \Bd(\Bk) \del_{k_i} \left( \frac{1}{\left| \Bd(\Bk) \right|} \right) \\
& = \frac{\del_{k_i} \Bd(\Bk)}{\big| \Bd(\Bk) \big|} - \frac{\Bd(\Bk)}{\big| \Bd(\Bk) \big|^2} \del_{k_i} \big| \Bd(\Bk) \big| \\
& = \frac{\del_{k_i} \Bd(\Bk)}{\big| \Bd(\Bk) \big|} - \frac{\Bd(\Bk)}{\big| \Bd(\Bk) \big|^2} \frac{\Bd(\Bk) \cdot \del_{k_i} \Bd(\Bk)}{\big| \Bd(\Bk) \big|} \\
& = \frac{1}{\big| \Bd(\Bk) \big|} \left\{ \del_{k_i} \Bd(\Bk) - \hat{\Bd}(\Bk) \left( \hat{\Bd}(\Bk) \cdot \del_{k_i} \Bd(\Bk) \right) \right\}\ . \\
}
Then we have that
\eqa{
[g_\pm(\Bk)]_{ii} = \frac{1}{4 \big| \Bd(\Bk) \big|^2} \bigg| \del_{k_i} \Bd(\Bk) - \hat{\Bd}(\Bk) \left( \hat{\Bd}(\Bk) \cdot \del_{k_i} \Bd(\Bk) \right) \bigg|^2\ .
\label{Pap_EQ:FSM in terms of d}
}
We want to find explicit expressions for both integrands close to the $\Gamma$ point. We compare two scenarios.

\subsubsection{$t' = 0$ \label{Pap_SSS: CBM t' = 0}}

In the first case, we set $t' = 0$. If $t' = 0$, then we have $\Bd(\Bk) = d_x(\Bk) \hat{\bsl{e}}_x$. We can expand $d_x(\Bk)$ in the region of the $\Gamma$ point, with $|\Bk| \ll 1$. We find
\eqa{
d_{x}(\Bk) = 4 t - \tfrac{t}{2}(k_x^2+k_y^2) + O(k^4)\ ,
\label{Pap_EQ:CB dx}
}
where $k=|\Bk|$. Then 
\eqa{
E_\pm(\Bk) & = \pm |d_x(\Bk)| \\
& = \pm \sqrt{d_x^2(\Bk)} \\
& = \pm \left[ 4 t - \frac{t}{2}(k_x^2 + k_y^2) + O(k^4) \right] \\
& = \pm d_x(\Bk) \\
}
to lowest order. Then $\Bnabla_\Bk E_\pm(\Bk) = \mp t \Bk$ at lowest order, which means $|\Bnabla_\Bk E_\pm(\Bk)| = t |\Bk|$. Thus, \cref{Pap_EQ:Lambda E FS Form} has the form
\eqa{
\la_E = \al_2 \int\limits_{\textsf{FS}_-} d\sSigma_{-,\Bk} \ t |\Bk| + O(k^2)\ ,
}
where
\eqa{
\al_d = \frac{\gamma^2}{m \langle \omega^2 \rangle} \frac{\sOmega}{(2\pi)^d}\ .
\label{Pap_EQ:Alpha d Definition}
}
We define the integral
\eqa{
\CI= \al_2\! \int\limits_{\textsf{FS}_-}\!\! d\sSigma_{-,\Bk} \>|\Bk| \quad\ ,
\label{Pap_EQ:CB_integral}
}
such that
\eqa{
\la_E = t \CI
}
to lowest order. We now examine $\la_{\rm geo}$. We can simplify \cref{Pap_EQ:FSM in terms of d}, with $|\Bd(\Bk)| = d_x(\Bk)$, and find
\eqa{
[g_\pm(\Bk)]_{ii} & = \frac{1}{4  d_x^2(\Bk)} \bigg| \del_{k_i} d_x(\Bk) \hat{\bsl{e}}_x - \frac{d_x(\Bk) \hat{\bsl{e}}_x}{d_x(\Bk)} \left( \frac{d_x(\Bk) \hat{\bsl{e}}_x}{d_x(\Bk)} \cdot \del_{k_i} d_x(\Bk) \hat{\bsl{e}}_x \right) \bigg|^2 \\
& = \frac{1}{4 d_x^2(\Bk)} \bigg| \del_{k_i} d_x(\Bk) \hat{\bsl{e}}_x - \frac{d_x^2(\Bk)}{d_x^2(\Bk)} \del_{k_i} d_x(\Bk) \hat{\bsl{e}}_x \bigg|^2 \\
& = \frac{1}{4 \big| d_x(\Bk) \big|^2} \bigg| \del_{k_i} d_x(\Bk) \left( 1 - \frac{d_x^2(\Bk)}{d_x^2(\Bk)} \right) \bigg|^2 \\
& = 0\ ,
}
since $1 - \frac{d_x^2(\Bk)}{d_x^2(\Bk)} = 0$. As such, $\la_{\rm geo} = 0$ to $O(k^2)$ near $\Gamma$.

\subsubsection{$t' \neq 0$ \label{Pap_SSS: CBM t' not = 0}}

We now allow $t'$ to take non-zero values. As such, we must consider the term $d_z(\Bk) = - 2 t' \left[ \cos(k_x) - \cos(k_y) \right]$. Thus, near $\Gamma$,
\eqa{
d_z(\Bk) = t' (k_x^2 - k_y^2) + O(k^4)\ .
\label{Pap_EQ:CB dz}
}
As such, $E_\pm(\Bk) = \pm \sqrt{d_x^2(\Bk) + d_z^2(\Bk)} = \pm d_x(\Bk) \sqrt{1 + \frac{d_z^2(\Bk)}{d_x^2(\Bk)}}$. From \cref{Pap_EQ:CB dz}, the leading order term is $O(k^2)$. However, the leading order term in \cref{Pap_EQ:CB dx} is $O(k^0)$. As such, $\frac{d_z^2(\Bk)}{d_x^2(\Bk)} \ll 1$, and we can say that $E_\pm = \pm d_x(\Bk)$. Thus, the dispersion remains the same to lowest order when $t' = 0$ and when $t' \neq 0$. As such, $\la_E$ will be identical to the case explored in \cref{Pap_SSS: CBM t' = 0}, as $\la_E$ is solely dependent on dispersion \cref{Pap_EQ:Lambda E FS Form}. So
\eqa{
\la_E = t \CI + O(k^2)\ ,
}
with $\CI$ defined in \cref{Pap_EQ:CB_integral}. We now return to $\la_{\rm geo}$. We now have $\Bd(\Bk) = d_x(\Bk) \hat{\bsl{e}}_x + d_z(\Bk) \hat{\bsl{e}}_z$. We know \cref{Pap_EQ:FSM in terms of d}. We have shown $E_\pm(\Bk) = \pm|\Bd(\Bk)| = \pm d_x(\Bk)$ to lowest order, as such $|\Bd(\Bk)| = d_x(\Bk)$. Then
\eqa{
\hat{\Bd}(\Bk) & = \frac{\Bd(\Bk)}{|\Bd(\Bk)|} \\
& = \frac{d_x(\Bk) \hat{\bsl{e}}_x + d_z(\Bk) \hat{\bsl{e}}_z}{d_x(\Bk)} \\
& = \hat{\bsl{e}}_x + \frac{d_z(\Bk)}{d_x(\Bk)} \hat{\bsl{e}}_z\ . \\
}
As previously discussed, $d_x(\Bk)$ is $O(k^0)$, while $d_z(\Bk)$ is $O(k^2)$. As such, $\frac{d_z(\Bk)}{d_x(\Bk)} \ll 1$, and $\hat{\Bd}(\Bk) = \hat{\bsl{e}}_x$ to lowest order in $k$. Substituting into \cref{Main:FSM in terms of d}, we have
\eqa{
[g_\pm(\Bk)]_{ii} & = \frac{1}{4 d_x^2(\Bk)} \bigg| \del_{k_i} \Bd(\Bk) - \hat{\bsl{e}}_x \left( \hat{\bsl{e}}_x \cdot \del_{k_i} \Bd(\Bk) \right) \bigg|^2 \\
& = \frac{1}{4 d_x^2(\Bk)} \bigg| \del_{k_i} \Bd(\Bk) - \hat{\bsl{e}}_x \left( \del_{k_i} d_x(\Bk) \right) \bigg|^2 \\
& = \frac{1}{4 d_x^2(\Bk)} \bigg| \del_{k_i} d_x(\Bk) \hat{\bsl{e}}_x + \del_{k_i} d_z(\Bk) \hat{\bsl{e}}_z - \del_{k_i} d_x(\Bk) \hat{\bsl{e}}_x \bigg|^2 \\
& = \frac{1}{4 d_x^2(\Bk)} \bigg| \del_{k_i} d_z(\Bk) \hat{\bsl{e}}_z \bigg|^2\ . \\
}
To lowest order, $\del_{k_x} d_z(\Bk) = 2 t' k_x$ and $\del_{k_y} d_z(\Bk) = -2 t' k_y$. Thus $| \del_{k_i} d_z(\Bk) \hat{\bsl{e}}_z |^2 = | 2 t' k_i |^2 = 4 t'^2 k_i^2$. Thus
\eqa{
\Tr [g_\pm(\Bk)] & = \sum_i \frac{4 t'^2 k_i^2}{4 d_x^2(\Bk)} \\
& = \frac{t'^2}{d_x^2(\Bk)} \lp k_x^2 + k_y^2 \rp \\
& = \frac{t'^2}{d_x^2(\Bk)} |\Bk|^2\ . \\
}
We also have $\Delta E(\Bk) = 2 |\Bd(\Bk)| = 2 d_x(\Bk)$, so $\Delta E^2(\Bk) = 4 d_x^2(\Bk)$. Since the dispersion remains the same, we have $|\Bnabla_\Bk E_\pm(\Bk)| = t |\Bk|$. So the integrand in \cref{Pap_EQ:Lambda Geo FS Form} is
\eqa{
\frac{\Delta E^2(\Bk)}{|\Bnabla_\Bk E_\pm(\Bk)|} \Tr [g_\pm(\Bk)] & = \frac{4 d_x^2(\Bk)}{t|\Bk|} \frac{t'^2}{d_x^2(\Bk)} |\Bk|^2 \\
& = \frac{4 t'^2}{t} |\Bk|\ . \\
}
As such, $\la_{\rm geo}$ has the form
\eqa{
\la_{\rm geo} = \frac{4 t'^2}{t} \CI + O(k^2)\ .
}
So, we have shown that the geometric contribution to $\la$ can be tuned while keeping the dispersive contribution constant, to lowest order in $|\Bk|$. We verify these results in \cref{Main_Fig:Geometric Enhancement}. We plot the exact integrals (\cref{Pap_EQ:Lambda E FS Form,Pap_EQ:Lambda Geo FS Form}) for $t' \in [0~\eV,t]$, compared with the integrals expanded to lowest order.

\section{The Hyper-diamond Lattice \label{Appendix: hyper-diamond Lattice}}

\subsection{Hyper-diamond Symmetry Properties \label{Pap_SS:HypD Symmetries}}

We now consider the generalized $d$ dimensional hyper-diamond model \cite{rachel_strain-induced_2016, kohler_nodal_2024}. Note that inversion is a symmetry operation in any dimension. In $d$ dimensions, each sublattice has $d+1$ neighbors, with nearest-neighbor vectors $\mbf{\hat{\delta}}_l^{(d)}$ where $l \in \{ 1, \dots, d+1 \}$. Each of these is a unit vector such that $\left|\mbf{\hat{\delta}}_l^{(d)} \right|^2 = 1$. In one dimension, we have $\mbf{\hat{\delta}}_1^{(1)} = \bsl{\hat{e}}_1$ and $\mbf{\hat{\delta}}_2^{(1)} = -\bsl{\hat{e}}_1$, where $\bsl{\hat{e}}_i$ is the unit vector in the $i$th cartesian direction. We may iteratively construct the $d$ dimensional hyper-diamond lattice
\eqa{
\hat{\mbf{\delta}}^{(d)}_l = \begin{cases}
    (1-d^{-2})^{1/2} \hat{\mbf{\delta}}^{(d-1)}_l - d^{-1} \hat{\bsl{e}}_d & \text{if } l \in \{1,\dots,d\} \\
    \hat{\bsl{e}}_d & \text{if } l = d+1 \\
\end{cases}\ ,
\label{Pap_EQ:hyper-diamond_Lattice}
}
where $\hat{\bsl{e}}_d$ is a unit vector in the $d$th cartesian direction. Furthermore, we will focus on the $d \geq 2$ case. Using the construction defined in \cref{Pap_EQ:hyper-diamond_Lattice}, it can be shown that
\eqa{
\hat{\mbf{\delta}}^{(d)}_l \cdot \hat{\mbf{\delta}}^{(d)}_{l'} = -d^{-1} + (1 + d^{-1}) \delta_{l l'}
\label{Pap_EQ:DeltaVecs_Orthogonality}
}
and
\eqa{
\sum_{l=1}^{d+1} \hat{\mbf{\delta}}^{(d)}_l = 0\ . \\
\label{Pap_EQ:DeltaVecs_Sums}
}
\cref{Pap_EQ:hyper-diamond_Lattice} can be used to define the primitive lattice vectors 
\eqa{
\Ba_{j}^{(d)} \equiv \hat{\mbf{\delta}}^{(d)}_{d+1} - \hat{\mbf{\delta}}^{(d)}_j = \hat{\bsl{e}}_d - \hat{\mbf{\delta}}^{(d)}_j\ .
\label{Pap_EQ:hyper-diamond PLV}
}
for $j \in \{1,\dots,d\}$. We take the origin to be between the two sublattices given by
\eqa{
\Btau_{\mu}^{(d)} & = \frac{(-1)^\mu}{2(d+1)} \sum_{i=1}^d \Ba_i^{(d)} \\
\label{Pap_EQ:tau_mu}
}
for $\mu = 1,2$. Using \cref{Pap_EQ:hyper-diamond PLV}, we can show that $\Btau_\mu^{(d)} = \frac{(-1)^\mu}{2} \hat{\mbf{\delta}}^{(d)}_{d+1} = \frac{(-1)^\mu}{2} \hat{\bsl{e}}_d$. The tight-binding Hamiltonian has the form
\begin{subequations}
\eqa{
h(\Bk) = t
\begin{pmatrix}
    0 & \gamma(\Bk) \\
    \gamma^*(\Bk) & 0
\end{pmatrix}
}
\eqa{
\gamma(\Bk) = \sum_{l=1}^{d+1} e^{-\ii \Bk \cdot \hat{\mbf{\delta}}^{(d)}_l}\ ,
}
\end{subequations}
where $t$ is a constant hopping parameter. For a two-band Hamiltonian of the form $h(\Bk) = \Bd(\Bk) \cdot \mbf{\sigma}$, we find $E_\pm(\Bk) = \pm | \Bd(\Bk) |$ and $P_\pm(\Bk) = \frac{1}{2} \sigma^0 \pm \frac{1}{2} \hat{\Bd}(\Bk) \cdot \mbf{\sigma}$, leading to \cref{Pap_EQ:FSM in terms of d}. As such, we can derive expressions for \cref{Pap_EQ:Lambda E FS Form,Pap_EQ:Lambda Geo FS Form}.

We now examine the symmetry properties of the $d$ dimensional hyper-diamond model, generated iteratively in \cref{Pap_EQ:hyper-diamond_Lattice}. We assume two spinless orbitals per unit cell, with orbital positions given by \cref{Pap_EQ:tau_mu}. Because $\Btau_1^{(d)} = -\Btau_2^{(d)}$, condition (ii) as described in \cref{Main:EPC_and_GA} is satisfied. In order to prove (i), we will show explicitly that the point group symmetries act on the sublattice as identity. We first define the vectors
\eqa{
& \mbf{\alpha}_i^{(d)} = \hat{\mbf{\delta}}^{(d)}_{i} - \hat{\mbf{\delta}}^{(d)}_{i+1} \\
& \mbf{\alpha}_{i+1}^{(d)} = \hat{\mbf{\delta}}^{(d)}_{i+1} - \hat{\mbf{\delta}}^{(d)}_{i+2} \\
\label{Pap_EQ:Alpha_eBasis}
}
for $i \in \{1,\dots,d-1\}$. In $d$-dimensions, we may show 
$\left|\mbf{\alpha}_i^{(d)}\right|^2 = 2(1+d^{-1})$. Consider a three-fold rotation about the 
plane $\Pi_i$, defined as $\Pi_i = \textsf{span } \big\{\mbf{\alpha}_i^{(d)},
\mbf{\alpha}_{i+1}^{(d)} \big\}$. The explicit form of an $n$-fold rotation $C_n$ about a plane spanned by two orthonormal vectors $\Bu^{(d)}, \Bv^{(d)} \in \dsR^d$ can be derived via 
the expression
\eqa{
C_n = I_d + \lp \cos(2\pi/n) - 1 \rp \lp \Bu^{(d)} \Bu^{(d)\ T} + \Bv^{(d)} \Bv^{(d)\ T}\rp + \sin(2\pi/n) \lp\Bv^{(d)} \Bu^{(d)\ T} - \Bu^{(d)} \Bv^{(d)\ T}\rp\ .
}
In the case of the hyper-diamond model, we may orthonormalize $\mbf{\alpha}_i^{(d)}, \mbf{\alpha}_{i+1}^{(d)}$, and find
\eqa{
& \Bu^{(d)} = \frac{\mbf{\alpha}_i^{(d)}}{| \mbf{\alpha}_i^{(d)} |} \\
& \Bv^{(d)} = \frac{\mbf{\alpha}_{i+1}^{(d)} - ( \mbf{\alpha}_{i+1}^{(d)} \cdot \Bu^{(d)}) \Bu^{(d)}}{| \mbf{\alpha}_{i+1}^{(d)} - ( \mbf{\alpha}_{i+1}^{(d)} \cdot \Bu^{(d)}) \Bu^{(d)} |}\ . \\
}
Inserting \cref{Pap_EQ:Alpha_eBasis}, we find
\eqa{
& \Bu^{(d)} = \frac{\hat{\mbf{\delta}}^{(d)}_{i} - \hat{\mbf{\delta}}^{(d)}_{i+1}}{\sqrt{2(1+d^{-1})}} \\
& \Bv^{(d)} = \frac{\hat{\mbf{\delta}}^{(d)}_{i} + \hat{\mbf{\delta}}^{(d)}_{i+1} - 2 \hat{\mbf{\delta}}^{(d)}_{i+2}}{\sqrt{6 (1+d^{-1})}} \ .
\label{Pap_EQ:uv_delta}
}
We may now derive the form of the three-fold rotation about the plane $\Pi_i$. First, we consider the case when $i = 1$. In the case of $i = 1$, we find that
\eqa{
C_{3}^{\Pi_1} = \begin{pmatrix}
    M_1 & 0 \\
    0 & I_{d-2} \\
\end{pmatrix}\ ,
\label{Pap_EQ:C3_1_Matrix}
}
where
\eqa{
M_1 =
\begin{pmatrix}
    -\frac{1}{2} & \frac{\sqrt{3}}{2} \\
    -\frac{\sqrt{3}}{2} & -\frac{1}{2} \\
\end{pmatrix}
}
and $I_{d-2}$ is the $d-2 \times d-2$ identity matrix. For $i \in \left\{ 2, \dots, d-1\right\}$, we have
\eqa{
C_{3}^{\Pi_i} = 
\begin{pmatrix}
    I_{i-2} & 0 & 0 \\
    0 & M_{i} & 0 \\
    0 & 0 & I_{d-(i+1)} \\ 
\end{pmatrix}\ ,
\label{Pap_EQ:C3i_inCartesian}
}
where $I_{i-2}$ is the $i-2 \times i-2$ identity matrix, $M_i$ is the $3 \times 3$ active rotation block, and $I_{d-(i+1)}$ is the $d-(i+1) \times d-(i+1)$ identity matrix. The active rotation block $M_i$ is given by
\eqa{
M_i = 
\begin{pmatrix}
\frac{1}{i} & \kappa \frac{\sqrt{\frac{i-1}{i+1}}}{i} & \kappa \sqrt{1 - \frac{2}{i(i+1)}} \\
\kappa \frac{\sqrt{(i-1)(i+1)}}{i} & - \frac{1}{i (i+1)} & - \frac{\sqrt{1 + \frac{2}{i}}}{i + 1} \\
0 & \frac{\sqrt{i (i+2)}}{i+1} & -\frac{1}{i+1} \\
\end{pmatrix}\ ,
}
where 
\eqa{
\kappa =
\begin{cases}
    -1 & \text{for } i = 2 \\
    1 & \text{otherwise} \\
\end{cases}\ .
}
So far, we have expressed $C_{3}^{\Pi_i}$ in the Cartesian basis $\hat{\bsl{e}}_j^{(d)}$. Since we will be acting these symmetry operations on the basis $\BR^{(d)} + \Btau_\mu^{(d)}$ which have simple forms in the non-orthonormal basis $\Ba_j^{(d)}$, it is convenient to express $C_{3}^{\Pi_i}$ in that basis. We denote the change of basis matrix as $\mathscr{A}_d = \mat{\Ba_1^{(d)} & \dots & \Ba_d^{(d)}}$. We first establish that $\mathscr{A}_d$ is invertible, which is required for the change of basis.

Note that the $k$th element of the column vector $\Ba_j^{(d)}$ is 0 if $k < j - 1$. As a result, $\mathscr{A}_d$ has the general form
\eqa{
\mathscr{A}_d = \begin{pmatrix}
    [\Ba_1^{(d)}]_1 & [\Ba_2^{(d)}]_1 & 0 & 0 & 0 &  \dots & 0 \\
    [\Ba_1^{(d)}]_2 & [\Ba_2^{(d)}]_2 & [\Ba_3^{(d)}]_2 & 0 & 0 & \dots & 0 \\
    [\Ba_1^{(d)}]_3 & [\Ba_2^{(d)}]_3 & [\Ba_3^{(d)}]_3 & [\Ba_4^{(d)}]_3 & 0 & \dots & 0 \\
    \vdots & \vdots & \vdots & \vdots & \vdots & & \vdots \\
    [\Ba_1^{(d)}]_d & [\Ba_2^{(d)}]_d & [\Ba_3^{(d)}]_d & [\Ba_4^{(d)}]_d & [\Ba_5^{(d)}]_d & \dots & [\Ba_d^{(d)}]_d \\
\end{pmatrix}\ .
}
Thus, $\mathscr{A}_d$ is an unreduced lower Hessenberg matrix \cite{golub_matrix_2013}, such that all elements $[\mathscr{A}_d]_{j,j+1} = [\Ba_{j+1}^{(d)}]_j \neq 0$ for $j \in \{1,\dots,d-1\}$. Consider the $j$th column of $\mathscr{A}_d$ for $j \in \{1,\dots,d-1\}$, and examine the $2 \times 2$ matrix
\eqa{
\begin{pmatrix}
    [\mathscr{A}_d]_{j,j} & [\mathscr{A}_d]_{j,j+1} \\
    [\mathscr{A}_d]_{j+1,j} & [\mathscr{A}_d]_{j+1,j+1} \\
\end{pmatrix}\ .
\label{Pap_EQ:Amat_subspace}
}
For every $j$, the entries in every column beyond $j+1$ are $0$. As such, we may construct an orthogonal rotation acting on the right which maps the row $\mat{[\mathscr{A}_d]_{j,j} & [\mathscr{A}_d]_{j,j+1} \\}$ to $\mat{r_j & 0}$, where $r_j = \sqrt{[\mathscr{A}_d]_{j,j}^2 + [\mathscr{A}_d]_{j,j+1}^2} \neq 0$. Let $G_j$ be an orthogonal rotation matrix whose only non-trivial $2 \times 2$ block acts on the subspace corresponding to \cref{Pap_EQ:Amat_subspace}. Then the matrix $\mathscr{A}_d' = \mathscr{A}_d \prod_{j=1}^{d-1} G_j$ is a lower-triangular matrix, as the rotation $G_{j+1}$ cannot reintroduce a non-zero element in the superdiagonal due to Hessenberg sparsity. Then 
\eqa{
\det \left[ \mathscr{A}_d' \right] & = \det \left[ \mathscr{A}_d \prod_{j=1}^{d-1} G_j \right] = \det \left[ \mathscr{A}_d \right] \det \left[ G_1 \right] \cdots \det \left[ G_{d-1} \right]\ . \\
}
Since $G_j$ are orthogonal rotation matrices, $\det \left[ G_j \right] = 1$. Thus $\det \left[ \mathscr{A}_d' \right] = \prod_{j} r_j = \det \left[ \mathscr{A}_d \right] \neq 0$, since $\mathscr{A}_d$ is unreduced. As such, we find that $\mathscr{A}_d$ is invertible in all dimensions $d$.

Having established invertibility, we proceed to evaluate $\wt{C}_{3}^{\Pi_i} = \mathscr{A}_d^{-1} C_{3}^{\Pi_i} \mathscr{A}_d$. It can be shown that, in the case $i < d - 1$,
\eqa{
\wt{C}_{3}^{\Pi_i} = 
\begin{pmatrix}
    I_{i-1} & 0 & 0 \\
    0 & P & 0 \\
    0 & 0 & I_{d-(i+1)} \\
\end{pmatrix}\ ,
\label{Pap_EQ:C3_in_aBasis}
}
where 
\eqa{
P = 
\begin{pmatrix}
    0 & 0 & 1 \\
    1 & 0 & 0 \\
    0 & 1 & 0 \\
\end{pmatrix}
}
is the cyclic permutation matrix. So, in the $\Ba_j^{(d)}$ basis, the rotation about the plane $\Pi_i$ simply permutes the coefficients of $\Ba_{i}^{(d)},\Ba_{i+1}^{(d)}$, and $\Ba_{i+2}^{(d)}$. If $i = d - 1$, however, $\wt{C}_{3}^{\Pi_i}$ takes a slightly more complicated form:
\eqa{
\wt{C}_{3}^{\Pi_{d-1}} = 
\begin{pmatrix}
    I_{d-2} & 0_{(d-2) \times 2} \\
    -\bsl{1} \\
    \Ba_{d-1}^{(d)}
\end{pmatrix}\ ,
}
where $-\bsl{1} = \mat{-1 & \dots & -1}$ is a row vector consisting of only $-1$ and $\Ba_{d-1}^{(d)} = \mat{0 & \dots & 0 & 1 & 0}$ is a row vector containing all zeros except for the $d-1$ element.

We now verify that the set of symmetry operations $C_{3}^{\Pi_i}$ preserve the sublattice. In order for the symmetry operation $\wt{C}_{3}^{\Pi_i}$ to preserve the sublattice, the operation must satisfy
\eqa{
\wt{C}_{3}^{\Pi_i} \Btau_\mu^{(d)} + \left( I_d - \wt{C}_{3}^{\Pi_i} \right) \Br_0^{(d)} = \Btau_\mu^{(d)} + \Delta \BR^{(d)}\ ,
\label{Pap_EQ:sub_pres}
}
where $\Delta \BR^{(d)}$ is a lattice vector. The rotation center $\Br_0^{(d)}$ is set to $\Btau_2^{(d)}$. As such, we may rewrite \cref{Pap_EQ:sub_pres} as
\eqa{
\Delta \BR^{(d)} & = \lp \wt{C}_{3}^{\Pi_i} - I_d \rp \Btau_\mu^{(d)} + \left( I_d - \wt{C}_{3}^{\Pi_i} \right) \Btau_2^{(d)} \\
& = \lp \wt{C}_{3}^{\Pi_i} - I_d \rp \lp \Btau_\mu^{(d)} - \Btau_2^{(d)} \rp \\
\label{Pap_EQ:sub_pres2}
}

We treat the cases $i < d - 1$ and $i = d - 1$ in turn. Recall \cref{Pap_EQ:tau_mu}. We apply \cref{Pap_EQ:C3_in_aBasis}, and find that $\wt{C}_{3}^{\Pi_i} \Btau_\mu^{(d)} = \Btau_\mu^{(d)}$ trivially, as $\wt{C}_{3}^{\Pi_i}$ is simply identity with a permutation block and all of the coefficients of $\Btau_\mu^{(d)}$ are identical in the $\Ba_j^{(d)}$ basis. If $\mu = 1$, we have
\eqa{
\Delta \BR^{(d)} & = \lp \wt{C}_{3}^{\Pi_i} - I_d \rp \lp \Btau_1^{(d)} - \Btau_2^{(d)} \rp \\
& = \lp \wt{C}_{3}^{\Pi_i} - I_d \rp \lp -\Btau_2^{(d)} - \Btau_2^{(d)} \rp \\
& = -2 \lp \wt{C}_{3}^{\Pi_i} - I_d \rp \Btau_2^{(d)} \\
& = 0\ . \\
}
If $\mu = 2$, we trivially find $\Delta\BR^{(d)} = 0$. As such, we find $\wt{C}_{3}^{\Pi_i}$ preserves the sublattice for $i < d - 1$. We now turn to the case $i = d - 1$. We again consider \cref{Pap_EQ:sub_pres2}. For $\mu = 1$, we have
\eqa{
\Delta \BR^{(d)} & = -2 \lp \wt{C}_{3}^{\Pi_{d-1}} - I_d \rp \Btau_2^{(d)} \\
& = -2 
\begin{pmatrix}
    0 & \dots \\
    \vdots & \\
    -1 & \dots & -1 & -2 & -1 \\
    0 & \dots & 0 & 1 & 0 \\
\end{pmatrix} \frac{1}{2(d+1)} \begin{pmatrix}
    1 \\ \vdots \\ 1 
\end{pmatrix} \\
& = \frac{1}{d+1} 
\begin{pmatrix}
    0 & \dots & d+1 & 0
\end{pmatrix}^T \\
& = \begin{pmatrix}
    0 & \dots & 1 & 0
\end{pmatrix}^T\ . \\
}
If $\mu = 2$, \cref{Pap_EQ:sub_pres2} is satisfied trivially with $\Delta\BR^{(d)} = 0$. As such, $\wt{C}_{3}^{\Pi_{d-1}}$ preserves the sublattice. Thus, we find $C_{3}^{\Pi_i}$ is a symmetry operation which preserves the sublattice for $i \in \left\{1, \dots, d-1 \right\}$.

We now use the symmetry operations $C_{3}^{\Pi_i}$ to solve the system of equations of the general form $W_{i,\Btau} = \sum_j R_{ji} W_{j,\Btau}$, introduced in \cref{Appendix: Introducing EPC}. We apply the three-fold rotations iteratively, beginning with $i = 1$. We find
\eqa{
\mat{
W_{1,\Btau} \\
W_{2,\Btau} \\
\vdots \\
W_{d,\Btau} \\
}
=
\mat{
-\frac{1}{2} & \frac{\sqrt{3}}{2} & 0 & \dots \\
-\frac{\sqrt{3}}{2} & -\frac{1}{2} & 0 & \dots \\
0 & 0 & 1 & \dots \\
\vdots & \vdots & \vdots & \ddots  \\
}
\mat{
W_{1,\Btau} \\
W_{2,\Btau} \\
\vdots \\
W_{d,\Btau} \\
}
=
\mat{
-\frac{1}{2} W_{1,\Btau} + \frac{\sqrt{3}}{2} W_{2,\Btau} \\
-\frac{\sqrt{3}}{2} W_{1,\Btau} - \frac{1}{2} W_{2,\Btau} \\
\vdots \\
W_{d,\Btau} \\
}\ .
\label{Pap_EQ:C1_on_Wi}
}
\cref{Pap_EQ:C1_on_Wi} generates a set of simultaneous equations:
\begin{subequations}
    \eqa{
W_{1,\Btau} = -\frac{1}{2} W_{1,\Btau} + \frac{\sqrt{3}}{2} W_{2,\Btau}
\label{Pap_EQ:W_1_EQ}
}
\eqa{
W_{2,\Btau} = -\frac{\sqrt{3}}{2} W_{1,\Btau} - \frac{1}{2} W_{2,\Btau}
\label{Pap_EQ:W_2_EQ}
}
\end{subequations}
Solving \cref{Pap_EQ:W_1_EQ} yields $W_{1,\Btau} = \frac{1}{\sqrt{3}} W_{2,\Btau}$. Then, plugging into \cref{Pap_EQ:W_2_EQ}, we find
\eqa{
W_{2,\Btau} & = -\frac{\sqrt{3}}{2} \left(  \frac{1}{\sqrt{3}} W_{2,\Btau} \right) - \frac{1}{2} W_{2,\Btau} \\
& = -\frac{1}{2} W_{2,\Btau} - \frac{1}{2} W_{2,\Btau} \\
& = -W_{2,\Btau}\ . \\
}
Thus, $W_{2,\Btau} = 0$ and $W_{1,\Btau} = 0$. In general, applying $C_{3}^{\Pi_i}$ to $W_{\Btau}$ gives relations $W_{i,\Btau} = \kappa \sqrt{\frac{i-1}{i+1}} W_{i-1,\Btau}$ and $W_{i+1,\Btau} = \kappa \sqrt{\frac{i(i-1)}{(i+1)(i+2)}} W_{i-1,\Btau}$ for $i \in \left\{2, \dots, d-1\right\}$. As $W_{1,\Btau} = W_{2,\Btau} = 0$, we find that $W_{i,\Btau} = 0$ for all $i \in \left\{1,\dots,d\right\}$. 

\subsection{Proof that $\frac{\lambda_{\rm geo}}{\lambda_E} = 1$ as $\mu \rightarrow 0^\pm$ in hyper-diamond \label{Sub Appendix: Proof of Ratio in d Dim}}

We now prove that $\frac{\lambda_{\rm geo}}{\lambda_E} = 1$ as $\mu \rightarrow 0^\pm$ in the hyper-diamond lattice in any dimension $d \geq 2$. Henceforth in this proof, we omit the superscript $(d)$. We begin by observing that 
\eqa{
\gamma(\Bk) = \sum_{l=1}^{d+1} e^{-\ii \Bk \cdot \hat{\mbf{\delta}}_l} = (1 + e^{\ii k_1} + \dots + e^{\ii k_d}) e^{-\ii \frac{(k_1 + \dots + k_d)}{d+1}}\ ,
}
if we take a general $\Bk$ to be a linear combination of primitive reciprocal lattice vectors,
\eqa{
\Bk = \frac{k_1}{2\pi}\Bb_1 + \dots + \frac{k_d}{2\pi}\Bb_d\ .
}
Suppose $\gamma(\Bp) = 0$, which means
\eq{
1 + e^{\ii p_1} + \dots + e^{\ii p_d} = 0\ .
}
We consider $\gamma(\Bp + \Bq)$ with $| \Bq | \ll 1$. We expand $\gamma(\Bk)$ with $\Bk \equiv \Bp + \Bq$:
\eqa{
\gamma(\Bk) = \sum_{l=1}^{d+1} e^{-\ii \Bk \cdot \hat{\mbf{\delta}}^{(d)}_l} & = \left(1 + \sum_{i=1,\dots,d}e^{\ii p_i + \ii q_i} \right) e^{-\ii \frac{\sum_{i=1,\dots,d}(p_i +q_i)}{d+1}} \\
& =  \left(1 + \sum_{i=1,\dots,d} e^{\ii p_i } ( 1 + \ii q_i) \right) e^{-\ii \frac{\sum_{i=1,\dots,d} p_i }{d+1}} \left[ 1-\ii \frac{\sum_{i=1,\dots,d} q_i}{d+1}\right] + O(q^2)\\
& =  \left(\sum_{i=1,\dots,d} e^{\ii p_i }  \ii q_i \right) e^{-\ii \theta_{\Bp}}  + O(q^2)\ ,
}
with $q = | \Bq |$ and $\theta_{\Bp} = e^{-\ii \frac{\sum_{i=1,\dots,d} p_i }{d+1}}$. We define
\eqa{
& \bsl{c}_{\Bp} = \sum_{i=1}^d \cos(p_i) \Ba_i \\
& \bsl{s}_{\Bp} = \sum_{i=1}^d \sin(p_i) \Ba_i\ ,
}
such that
\eqa{
\gamma(\Bp + \Bq) & = (\ii \bsl{c}_{\Bp} \cdot  \Bq - \bsl{s}_{\Bp} \cdot  \Bq) e^{-\ii \theta_{\Bp}}  + O(q^2)\ ,
}
leading to
\eq{
h(\Bp+\Bq) = \mat{ & (\ii \bsl{c}_{\Bp} \cdot  \Bq - \bsl{s}_{\Bp} \cdot  \Bq) e^{-\ii \theta_{\Bp}}  \\
(-\ii \bsl{c}_{\Bp} \cdot  \Bq - \bsl{s}_{\Bp} \cdot  \Bq) e^{\ii \theta_{\Bp}} & }
+ O(q^2)\ ,
}
where we have set $t=1$ WLOG. From this, we find the dispersion reads
\eqa{
E_{\pm,\Bp+\Bq} = \pm \sqrt{ (\bsl{c}_{\Bp} \cdot  \Bq)^2 + (\bsl{s}_{\Bp} \cdot  \Bq)^2} + O(q^2)\ ,
}
and the projector for $E_{-,\Bp+\Bq}$ reads
\eq{
P_{-,\Bp+\Bq} = \frac{1}{2} - \frac{1}{2} \mat{ & e^{-\ii \theta_{\Bp}-\ii \phi_{\Bp,\Bq}}  \\
 e^{\ii \theta_{\Bp}+\ii \phi_{\Bp,\Bq}} & } + O(q^1)\ ,
}
where
\eq{
e^{-\ii \phi_{\Bp,\Bq}} = \frac{\ii \bsl{c}_{\Bp} \cdot  \Bq - \bsl{s}_{\Bp} \cdot  \Bq}{\sqrt{ (\bsl{c}_{\Bp} \cdot  \Bq)^2 + (\bsl{s}_{\Bp} \cdot  \Bq)^2}}\ .
}

From now on, we only take leading order. Then, we know
\eqa{
& \partial_{k_i} E_{-,\Bp+\Bq} = \partial_{q_i} E_{-,\Bp+\Bq} = \frac{ \cos(p_i) (\bsl{c}_{\Bp} \cdot  \Bq) + \sin(p_i) (\bsl{s}_{\Bp} \cdot  \Bq) }{E_{-,\Bp+\Bq}}\ .
}

To obtain $|\Bnabla_{\Bk} E_{-,\Bp+\Bq}|$, we note that
\eq{
\frac{\partial}{\partial \Bk\cdot\hat{\bsl{e}}_i} = \sum_j \frac{\partial k_j}{\partial \Bk\cdot\hat{\bsl{e}}_i} \frac{\partial}{\partial k_j } = \sum_j \frac{\partial \Bk\cdot\Ba_j}{\partial \Bk\cdot\hat{\bsl{e}}_i} = \sum_{j} A_{ji}\frac{\partial}{\partial k_j }
}
with
\eq{
A_{ji} =\Ba_j\cdot \hat{\bsl{e}}_i\ ,
}
leading to
\eq{
\Bnabla_{\Bk} = \sum_{i} \hat{\bsl{e}}_i \frac{\partial}{\partial \Bk\cdot\hat{\bsl{e}}_i} =\sum_{j}  \sum_{i} \hat{\bsl{e}}_i A_{ji} \frac{\partial}{\partial k_j } =\sum_{j}  \sum_{i} \hat{\bsl{e}}_i (\Ba_j\cdot \hat{\bsl{e}}_i) \frac{\partial}{\partial k_j }  =\sum_{j}  \Ba_j \frac{\partial}{\partial k_j }\ .
}

Thus,
\eqa{
\Bnabla_{\Bk} E_{-,\Bp+\Bq} = \Bnabla_{\Bq}E_{-,\Bp+\Bq} = \sum_{i} \Ba_i \frac{\cos(p_i) (\bsl{c}_{\Bp} \cdot  \Bq) + \sin(p_i) (\bsl{s}_{\Bp} \cdot  \Bq) }{E_{-,\Bp+\Bq}} =  \frac{\bsl{c}_{\Bp}  (\bsl{c}_{\Bp} \cdot  \Bq) + \bsl{s}_{\Bp} (\bsl{s}_{\Bp} \cdot  \Bq) }{E_{-,\Bp+\Bq}}\ ,
} 
leading to
\eq{
|\Bnabla_{\Bk} E_{-,\Bp+\Bq}| = \frac{|\bsl{c}_{\Bp}  (\bsl{c}_{\Bp} \cdot  \Bq) + \bsl{s}_{\Bp} (\bsl{s}_{\Bp} \cdot  \Bq)| }{\sqrt{ (\bsl{c}_{\Bp} \cdot  \Bq)^2 + (\bsl{s}_{\Bp} \cdot  \Bq)^2}}\ .
}
Similarly,
\eqa{
\frac{1}{2}\Tr[\partial_{k_i} P_{-,\Bp+\Bq} \partial_{k_j} P_{-,\Bp+\Bq} ] & = \frac{1}{2}\Tr[\partial_{q_i} P_{-,\Bp+\Bq} \partial_{q_j} P_{-,\Bp+\Bq} ] \\
& = \frac{1}{8} \Tr\left[\mat{ & \partial_{q_i} e^{-\ii \theta_{\Bp}-\ii \phi_{\Bp,\Bq}}  \\
 \partial_{q_i} e^{\ii \theta_{\Bp}+\ii \phi_{\Bp,\Bq}} & } \mat{ & \partial_{q_j}  e^{-\ii \theta_{\Bp}-\ii \phi_{\Bp,\Bq}}  \\
\partial_{q_j}  e^{\ii \theta_{\Bp}+\ii \phi_{\Bp,\Bq}} & }\right] \\
& = \frac{1}{8} (\partial_{q_i} e^{-\ii \phi_{\Bp,\Bq}} \partial_{q_j}  e^{\ii \phi_{\Bp,\Bq}} + \partial_{q_i} e^{\ii \phi_{\Bp,\Bq}} \partial_{q_j}  e^{-\ii \phi_{\Bp,\Bq}} )\ . \\
}
Combined with
\eqa{
\partial_{q_i} e^{-\ii \phi_{\Bp,\Bq}} & = \frac{\ii \cos(p_i) - \sin(p_i)}{\sqrt{ (\bsl{c}_{\Bp} \cdot  \Bq)^2 + (\bsl{s}_{\Bp} \cdot  \Bq)^2}} - \frac{\ii \bsl{c}_{\Bp} \cdot  \Bq - \bsl{s}_{\Bp} \cdot  \Bq}{ \sqrt{ (\bsl{c}_{\Bp} \cdot  \Bq)^2 + (\bsl{s}_{\Bp} \cdot  \Bq)^2}} \frac{\cos(p_i) (\bsl{c}_{\Bp} \cdot  \Bq) + \sin(p_i) (\bsl{s}_{\Bp} \cdot  \Bq)}{(\bsl{c}_{\Bp} \cdot  \Bq)^2 + (\bsl{s}_{\Bp} \cdot  \Bq)^2} \\
&  = \frac{\ii \cos(p_i) - \sin(p_i)}{\sqrt{ (\bsl{c}_{\Bp} \cdot  \Bq)^2 + (\bsl{s}_{\Bp} \cdot  \Bq)^2}} \frac{(\bsl{c}_{\Bp} \cdot  \Bq)^2 + (\bsl{s}_{\Bp} \cdot  \Bq)^2}{(\bsl{c}_{\Bp} \cdot  \Bq)^2 + (\bsl{s}_{\Bp} \cdot  \Bq)^2} - \frac{\ii \bsl{c}_{\Bp} \cdot  \Bq - \bsl{s}_{\Bp} \cdot  \Bq}{ \sqrt{ (\bsl{c}_{\Bp} \cdot  \Bq)^2 + (\bsl{s}_{\Bp} \cdot  \Bq)^2}} \frac{\cos(p_i) (\bsl{c}_{\Bp} \cdot  \Bq) + \sin(p_i) (\bsl{s}_{\Bp} \cdot  \Bq)}{(\bsl{c}_{\Bp} \cdot  \Bq)^2 + (\bsl{s}_{\Bp} \cdot  \Bq)^2} \\
&  = \frac{\ii \cos(p_i) (\bsl{c}_{\Bp} \cdot  \Bq)^2 - \sin(p_i) (\bsl{c}_{\Bp} \cdot  \Bq)^2 + \ii \cos(p_i) (\bsl{s}_{\Bp} \cdot  \Bq)^2  - \sin(p_i) (\bsl{s}_{\Bp} \cdot  \Bq)^2}{[ (\bsl{c}_{\Bp} \cdot  \Bq)^2 + (\bsl{s}_{\Bp} \cdot  \Bq)^2]^{3/2}} \\
& \qquad - \frac{\ii \cos(p_i) (\bsl{c}_{\Bp} \cdot  \Bq)^2 - \cos(p_i) (\bsl{c}_{\Bp} \cdot  \Bq)(\bsl{s}_{\Bp} \cdot  \Bq) + \ii \sin(p_i) (\bsl{s}_{\Bp} \cdot  \Bq) (\bsl{c}_{\Bp} \cdot  \Bq) - \sin(p_i) (\bsl{s}_{\Bp} \cdot  \Bq)^2}{[ (\bsl{c}_{\Bp} \cdot  \Bq)^2 + (\bsl{s}_{\Bp} \cdot  \Bq)^2]^{3/2}} \\
&  = \frac{ - \sin(p_i) (\bsl{c}_{\Bp} \cdot  \Bq)^2 + \ii \cos(p_i) (\bsl{s}_{\Bp} \cdot  \Bq)^2 + \cos(p_i) (\bsl{c}_{\Bp} \cdot  \Bq)(\bsl{s}_{\Bp} \cdot  \Bq) - \ii \sin(p_i) (\bsl{s}_{\Bp} \cdot  \Bq) (\bsl{c}_{\Bp} \cdot  \Bq)}{[ (\bsl{c}_{\Bp} \cdot  \Bq)^2 + (\bsl{s}_{\Bp} \cdot  \Bq)^2]^{3/2}} \\
&  = \frac{ [ \cos(p_i) (\bsl{s}_{\Bp} \cdot  \Bq) - \sin(p_i) (\bsl{c}_{\Bp} \cdot  \Bq) ]  (\bsl{c}_{\Bp} \cdot  \Bq + \ii \bsl{s}_{\Bp}\cdot\Bq)}{[ (\bsl{c}_{\Bp} \cdot  \Bq)^2 + (\bsl{s}_{\Bp} \cdot  \Bq)^2]^{3/2}} \\
&  = \frac{  \cos(p_i) (\bsl{s}_{\Bp} \cdot  \Bq) - \sin(p_i) (\bsl{c}_{\Bp} \cdot  \Bq) }{(\bsl{c}_{\Bp} \cdot  \Bq)^2 + (\bsl{s}_{\Bp} \cdot  \Bq)^2} (-\ii e^{-\ii \phi_{\Bp,\Bq}})\ ,
}
which means
\eqa{
\partial_{q_i} e^{-\ii \phi_{\Bp,\Bq}} \partial_{q_j}  e^{\ii \phi_{\Bp,\Bq}}  = \frac{  \cos(p_i) (\bsl{s}_{\Bp} \cdot  \Bq) - \sin(p_i) (\bsl{c}_{\Bp} \cdot  \Bq) }{(\bsl{c}_{\Bp} \cdot  \Bq)^2 + (\bsl{s}_{\Bp} \cdot  \Bq)^2} \frac{  \cos(p_j) (\bsl{s}_{\Bp} \cdot  \Bq) - \sin(p_j) (\bsl{c}_{\Bp} \cdot  \Bq) }{(\bsl{c}_{\Bp} \cdot  \Bq)^2 + (\bsl{s}_{\Bp} \cdot  \Bq)^2}\ .
}

It leads to
\eqa{
\frac{1}{2}\Tr[\partial_{k_i} P_{-,\Bp+\Bq} \partial_{k_j} P_{-,\Bp+\Bq} ] & = \frac{1}{4}  \frac{  \cos(p_i) (\bsl{s}_{\Bp} \cdot  \Bq) - \sin(p_i) (\bsl{c}_{\Bp} \cdot  \Bq) }{(\bsl{c}_{\Bp} \cdot  \Bq)^2 + (\bsl{s}_{\Bp} \cdot  \Bq)^2} \frac{  \cos(p_j) (\bsl{s}_{\Bp} \cdot  \Bq) - \sin(p_j) (\bsl{c}_{\Bp} \cdot  \Bq) }{(\bsl{c}_{\Bp} \cdot  \Bq)^2 + (\bsl{s}_{\Bp} \cdot  \Bq)^2}\ . \\
}
Note that
\eqa{
\Tr[g(\Bk)] & = \sum_i\frac{1}{2}\Tr[\partial_{\Bk\cdot \hat{\bsl{e}}_i} P_{-,\Bp+\Bq} \partial_{\Bk\cdot \hat{\bsl{e}}_i} P_{-,\Bp+\Bq} ] = \sum_i \sum_{j_1,j_2} A_{j_1 i } A_{j_2 i} \frac{1}{2}\Tr[\partial_{k_{j_1}} P_{-,\Bp+\Bq} \partial_{k_{j_2}} P_{-,\Bp+\Bq} ] \\
& = \frac{1}{4}  \frac{  \bsl{c}_{\Bp} (\bsl{s}_{\Bp} \cdot  \Bq) - \bsl{s}_{\Bp} (\bsl{c}_{\Bp} \cdot  \Bq) }{(\bsl{c}_{\Bp} \cdot  \Bq)^2 + (\bsl{s}_{\Bp} \cdot  \Bq)^2} \cdot \frac{  \bsl{c}_{\Bp} (\bsl{s}_{\Bp} \cdot  \Bq) -  \bsl{s}_{\Bp} (\bsl{c}_{\Bp} \cdot  \Bq) }{(\bsl{c}_{\Bp} \cdot  \Bq)^2 + (\bsl{s}_{\Bp} \cdot  \Bq)^2}\ ,
}
which gives
\eqa{
\frac{\Delta E^2(\Bk)}{|\Bnabla_{\Bk} E_{-}(\Bk)|}\Tr[g(\Bk)] =  \frac{  \bsl{c}_{\Bp} (\bsl{s}_{\Bp} \cdot  \Bq) - \bsl{s}_{\Bp} (\bsl{c}_{\Bp} \cdot  \Bq) }{|\bsl{c}_{\Bp}  (\bsl{c}_{\Bp} \cdot  \Bq) + \bsl{s}_{\Bp} (\bsl{s}_{\Bp} \cdot  \Bq)|} \cdot \frac{  \bsl{c}_{\Bp} (\bsl{s}_{\Bp} \cdot  \Bq) -  \bsl{s}_{\Bp} (\bsl{c}_{\Bp} \cdot  \Bq) }{[(\bsl{c}_{\Bp} \cdot  \Bq)^2 + (\bsl{s}_{\Bp} \cdot  \Bq)^{2}]^{1/2}}\ .
}

For $\mu \rightarrow 0^-$, points $\Bk$ that lie on the Fermi surface satisfy $|\Bq| \ll 1$. Moreover, in this case, 
\eqa{
[(\bsl{c}_{\Bp} \cdot  \Bq)^2 + (\bsl{s}_{\Bp} \cdot  \Bq)^{2}]^{1/2} = |\mu|\ .
}
Then, we know
\eqa{
\frac{\Delta E^2(\Bk)}{|\Bnabla_{\Bk} E_{-}(\Bk)|}\Tr[g(\Bk)] =  \frac{  |\bsl{c}_{\Bp} (\bsl{s}_{\Bp} \cdot  \Bq) - \bsl{s}_{\Bp} (\bsl{c}_{\Bp} \cdot  \Bq)|^2 }{|\bsl{c}_{\Bp}  (\bsl{c}_{\Bp} \cdot  \Bq) + \bsl{s}_{\Bp} (\bsl{s}_{\Bp} \cdot  \Bq)|}\frac{1}{ |\mu|}
}
and
\eq{
|\Bnabla_{\Bk} E_{-,\Bp+\Bq}| = \frac{1}{ |\mu|}|\bsl{c}_{\Bp}  (\bsl{c}_{\Bp} \cdot  \Bq) + \bsl{s}_{\Bp} (\bsl{s}_{\Bp} \cdot  \Bq)|\ .
}

Now we integrate over the Fermi surface. For small $| \mu |$, we may split the Fermi surface. In $d$ dimensions, $\Bp$ forms a $d-2$ dimensional manifold. This manifold is the Fermi surface defined by $\mu = 0$, $\textsf{FS}_0$. We define $dV_\Bp$, the $d-2$ dimensional volume element at $\Bp$. For a fixed $\Bp$, $\Bp + \bsl{c}_{\Bp}$ and $\Bp + \bsl{s}_{\Bp}$ are perpendicular to $\textsf{FS}_0$ at $\Bp$, since a non-zero $\Bq$ will change the energy. In other words, we can define a plane generated by $\Bp + \bsl{c}_{\Bp}$ and $\Bp + \bsl{s}_{\Bp}$ that is perpendicular to $\textsf{FS}_0$. We now restrict $\Bp+\Bq$ to this 2D plane. We require $\Bq \in \textsf{FS}$, which is equivalent to requiring $\bsl{c}_{\Bp}\cdot\Bq = |\mu| \cos(\theta)$ and $\bsl{s}_{\Bp}\cdot\Bq = |\mu| \sin(\theta)$. In other words, we require
\eqa{
\Bq = |\mu|  \cos(\theta) \bsl{c}_{\Bp}^*  + |\mu|  \sin(\theta) \bsl{s}_{\Bp}^*\ ,
}
where
\eqa{
\bsl{c}_{\Bp}^* = \frac{|\bsl{s}_{\Bp}|^2 \bsl{c}_{\Bp} - |\bsl{c}_{\Bp}\cdot \bsl{s}_{\Bp}| \bsl{s}_{\Bp}}{|\bsl{c}_{\Bp}|^2|\bsl{s}_{\Bp}|^2 - |\bsl{c}_{\Bp}\cdot \bsl{s}_{\Bp}|^2}\ , \  \bsl{s}_{\Bp}^* = \frac{|\bsl{c}_{\Bp}|^2 \bsl{s}_{\Bp} - |\bsl{c}_{\Bp}\cdot \bsl{s}_{\Bp}| \bsl{c}_{\Bp}}{|\bsl{c}_{\Bp}|^2|\bsl{s}_{\Bp}|^2 - |\bsl{c}_{\Bp}\cdot \bsl{s}_{\Bp}|^2}\ .
}
This means that
\eqa{
\left|\frac{\partial \Bq}{\partial \theta}\right| & = |\mu|  | - \sin(\theta) \bsl{c}_{\Bp}^*  +  \cos(\theta) \bsl{s}_{\Bp}^*| = |\mu|\sqrt{\sin^2(\theta) (\bsl{c}_{\Bp}^*)^2 + \cos^2(\theta) (\bsl{s}_{\Bp}^*)^2 - 2 \cos(\theta) \sin(\theta) \bsl{c}_{\Bp}^*\cdot \bsl{s}_{\Bp}^* } \\
& = \frac{|\mu| \sqrt{ \sin^2(\theta) (|\bsl{s}_{\Bp}|^2 \bsl{c}_{\Bp} - |\bsl{c}_{\Bp}\cdot \bsl{s}_{\Bp}| \bsl{s}_{\Bp})^2  + \cos^2(\theta) (|\bsl{c}_{\Bp}|^2 \bsl{s}_{\Bp} - |\bsl{c}_{\Bp}\cdot \bsl{s}_{\Bp}| \bsl{c}_{\Bp})^2 +  2 \cos(\theta) \sin(\theta) |\bsl{c}_{\Bp}\cdot \bsl{s}_{\Bp}|(|\bsl{s}_{\Bp}|^2 \bsl{c}_{\Bp}^2  - |\bsl{c}_{\Bp}\cdot \bsl{s}_{\Bp}|^2 )} }{|\bsl{c}_{\Bp}|^2|\bsl{s}_{\Bp}|^2 - |\bsl{c}_{\Bp}\cdot \bsl{s}_{\Bp}|^2}\\
& = \frac{|\mu| \sqrt{ \sin^2(\theta) (|\bsl{s}_{\Bp}|^4 \bsl{c}_{\Bp}^2 - |\bsl{c}_{\Bp}\cdot \bsl{s}_{\Bp}|^2 \bsl{s}_{\Bp}^2 )  + \cos^2(\theta) (|\bsl{c}_{\Bp}|^4 \bsl{s}_{\Bp}^2 - |\bsl{c}_{\Bp}\cdot \bsl{s}_{\Bp}|^2 \bsl{c}_{\Bp}^2 ) + 2 \cos(\theta) \sin(\theta) |\bsl{c}_{\Bp}\cdot \bsl{s}_{\Bp}|(|\bsl{s}_{\Bp}|^2 \bsl{c}_{\Bp}^2  - |\bsl{c}_{\Bp}\cdot \bsl{s}_{\Bp}|^2 ) } } {|\bsl{c}_{\Bp}|^2|\bsl{s}_{\Bp}|^2 - |\bsl{c}_{\Bp}\cdot \bsl{s}_{\Bp}|^2}\\
& = \frac{|\mu| \sqrt{ \sin^2(\theta) \bsl{s}_{\Bp}^2(|\bsl{s}_{\Bp}|^2 \bsl{c}_{\Bp}^2 - |\bsl{c}_{\Bp}\cdot \bsl{s}_{\Bp}|^2  )  + \bsl{c}_{\Bp}^2  \cos^2(\theta) (|\bsl{c}_{\Bp}|^2 \bsl{s}_{\Bp}^2 - |\bsl{c}_{\Bp}\cdot \bsl{s}_{\Bp}|^2 )+ 2 \cos(\theta) \sin(\theta) |\bsl{c}_{\Bp}\cdot \bsl{s}_{\Bp}|(|\bsl{s}_{\Bp}|^2 \bsl{c}_{\Bp}^2  - |\bsl{c}_{\Bp}\cdot \bsl{s}_{\Bp}|^2 ) } }{|\bsl{c}_{\Bp}|^2|\bsl{s}_{\Bp}|^2 - |\bsl{c}_{\Bp}\cdot \bsl{s}_{\Bp}|^2}\\
& = \frac{|\mu| \sqrt{ \sin^2(\theta) \bsl{s}_{\Bp}^2  + \bsl{c}_{\Bp}^2  \cos^2(\theta) + 2 \cos(\theta) \sin(\theta) |\bsl{c}_{\Bp}\cdot \bsl{s}_{\Bp}|} }{\sqrt{|\bsl{c}_{\Bp}|^2|\bsl{s}_{\Bp}|^2 - |\bsl{c}_{\Bp}\cdot \bsl{s}_{\Bp}|^2}} \\
& = \frac{|\mu| |  \sin(\theta) \bsl{s}_{\Bp} +  \bsl{c}_{\Bp} \cos(\theta)|}{\sqrt{|\bsl{c}_{\Bp}|^2|\bsl{s}_{\Bp}|^2 - |\bsl{c}_{\Bp}\cdot \bsl{s}_{\Bp}|^2}} = \frac{|  (\Bq\cdot\bsl{s}_{\Bp}) \bsl{s}_{\Bp} +  \bsl{c}_{\Bp} (\Bq\cdot\bsl{c}_{\Bp}) |}{\sqrt{|\bsl{c}_{\Bp}|^2|\bsl{s}_{\Bp}|^2 - |\bsl{c}_{\Bp}\cdot \bsl{s}_{\Bp}|^2}}\ .
}
Then, we know the surface element on the $\textsf{FS}$ is
\eqa{
d\sSigma = d V_{\Bp} d\theta \left|\frac{\partial \Bq}{\partial \theta}\right| = d V_{\Bp} d\theta \frac{|\mu| |  (\Bq\cdot\bsl{s}_{\Bp}) \bsl{s}_{\Bp} +  \bsl{c}_{\Bp} (\Bq\cdot\bsl{c}_{\Bp}) |}{\sqrt{|\bsl{c}_{\Bp}|^2|\bsl{s}_{\Bp}|^2 - |\bsl{c}_{\Bp}\cdot \bsl{s}_{\Bp}|^2}}\ .
}
Then, using \cref{Pap_EQ:Lambda Geo FS Form,Pap_EQ:Lambda E FS Form}, we find
\eqa{
\int\limits_{\textsf{FS}}d\sSigma  |\Bnabla_{\Bk} E_{-,\Bp+\Bq}| & = \int\limits_{\textsf{FS}_0} d V_{\Bp} \int_0^{2\pi} d\theta \frac{|\mu| |  (\Bq\cdot\bsl{s}_{\Bp}) \bsl{s}_{\Bp} +  \bsl{c}_{\Bp} (\Bq\cdot\bsl{c}_{\Bp}) |}{\sqrt{|\bsl{c}_{\Bp}|^2|\bsl{s}_{\Bp}|^2 - |\bsl{c}_{\Bp}\cdot \bsl{s}_{\Bp}|^2}}  \frac{1}{ |\mu|}|\bsl{c}_{\Bp}  (\bsl{c}_{\Bp} \cdot  \Bq) + \bsl{s}_{\Bp} (\bsl{s}_{\Bp} \cdot  \Bq)| \\
&  =  \int\limits_{\textsf{FS}_0} d V_{\Bp} \int_0^{2\pi} d\theta \frac{ | \sin(\theta) \bsl{s}_{\Bp} +  \bsl{c}_{\Bp} \cos(\theta) |^2}{\sqrt{|\bsl{c}_{\Bp}|^2|\bsl{s}_{\Bp}|^2 - |\bsl{c}_{\Bp}\cdot \bsl{s}_{\Bp}|^2}} \\
&  =  \int\limits_{\textsf{FS}_0} d V_{\Bp} \frac{ \pi ( \bsl{s}_{\Bp}^2 + \bsl{c}_{\Bp}^2)}{\sqrt{|\bsl{c}_{\Bp}|^2|\bsl{s}_{\Bp}|^2 - |\bsl{c}_{\Bp}\cdot \bsl{s}_{\Bp}|^2}} 
}

\eqa{
\int\limits_{\textsf{FS}}d\sSigma  \frac{\Delta E^2(\Bk)}{|\Bnabla_{\Bk} E_{-}(\Bk)|}\Tr[g(\Bk)]  & = \int\limits_{\textsf{FS}_0} d V_{\Bp} \int_0^{2\pi} d\theta \frac{|\mu| |  (\Bq\cdot\bsl{s}_{\Bp}) \bsl{s}_{\Bp} +  \bsl{c}_{\Bp} (\Bq\cdot\bsl{c}_{\Bp}) |}{\sqrt{|\bsl{c}_{\Bp}|^2|\bsl{s}_{\Bp}|^2 - |\bsl{c}_{\Bp}\cdot \bsl{s}_{\Bp}|^2}} \frac{  |\bsl{c}_{\Bp} (\bsl{s}_{\Bp} \cdot  \Bq) - \bsl{s}_{\Bp} (\bsl{c}_{\Bp} \cdot  \Bq)|^2 }{|\bsl{c}_{\Bp}  (\bsl{c}_{\Bp} \cdot  \Bq) + \bsl{s}_{\Bp} (\bsl{s}_{\Bp} \cdot  \Bq)|}\frac{1}{ |\mu|} \\
& = \int\limits_{\textsf{FS}_0} d V_{\Bp} \int_0^{2\pi} d\theta \frac{|\bsl{c}_{\Bp} (\bsl{s}_{\Bp} \cdot  \Bq) - \bsl{s}_{\Bp} (\bsl{c}_{\Bp} \cdot  \Bq)|^2}{\sqrt{|\bsl{c}_{\Bp}|^2|\bsl{s}_{\Bp}|^2 - |\bsl{c}_{\Bp}\cdot \bsl{s}_{\Bp}|^2}} \\
& = \int\limits_{\textsf{FS}_0} d V_{\Bp} \int_0^{2\pi} d\theta \frac{|\bsl{c}_{\Bp} \sin(\theta) - \bsl{s}_{\Bp} \cos(\theta)|^2}{\sqrt{|\bsl{c}_{\Bp}|^2|\bsl{s}_{\Bp}|^2 - |\bsl{c}_{\Bp}\cdot \bsl{s}_{\Bp}|^2}} \\
& = \int\limits_{\textsf{FS}_0} d V_{\Bp}  \frac{\pi(\bsl{c}_{\Bp}^2 + \bsl{s}_{\Bp}^2 )}{\sqrt{|\bsl{c}_{\Bp}|^2|\bsl{s}_{\Bp}|^2 - |\bsl{c}_{\Bp}\cdot \bsl{s}_{\Bp}|^2}}\ .
}
Thus, $\frac{\la_{\rm geo}}{\la_E} = 1$, and $\frac{\la_{\rm geo}}{\la_E + \la_{\rm geo}} = \frac{1}{2}$ as $\mu \rightarrow 0^-$. A similar result can be performed for $\mu \rightarrow 0^+$, which yields the same conclusion.

\end{document}